\documentclass[aps,twocolumn,letterpaper,preprintnumbers,amsmath,amssymb,floatfix,nofootinbib]{revtex4}
\usepackage{graphicx}
\usepackage{bm}
\usepackage{color}
\usepackage{hyperref}
\hypersetup{
	colorlinks=true,
	linkcolor=blue,
	urlcolor=blue,
	citecolor=blue,
	pdfpagemode=FullScreen,
}

\def\nablab{{\bm \nabla}}

\def\O{\mathcal{O}}

\usepackage{color}
\definecolor{gray}{rgb}{0.5,0.5,0.5}
\definecolor{lgray}{rgb}{0.8,0.8,0.8}
\definecolor{dgray}{rgb}{0.6,0.6,0.6}
\definecolor{dred}{rgb}{0.5,0.0,0.0}
\definecolor{dgreen}{rgb}{0.0,0.5,0.0}
\definecolor{dblue}{rgb}{0.0,0.0,0.5}
\definecolor{violet}{rgb}{0.7,0.0,0.5}

\def\L{{\mathcal L}}

\def\O{{\mathcal O}}

\def\E{{\mathcal E}}

\def\rpeak{r_{\rm peak}}

\def\figures{.}

\begin{document}

\preprint{}

\title{Frequency-inference method for reduced modeling of energetic particle modes (EPM) utilizing resonant auto-optimization remnants of imperfect time-scale separation\vspace{-0.25cm}}

\author{Andreas~Bierwage$^{1}$\footnote{e-mail:bierwage.andreas@qst.go.jp}, Vin\'{i}cius~N.~Duarte$^{2}$, Pablo~Oyola$^{2}$, Kouji~Shinohara$^{3,1}$, Roscoe~B.~White$^{2}$}

\affiliation{
$^{1}$ National Institutes for Quantum Science and Technology (QST), Naka Institute for Fusion Science and Technology, Ibaraki 311-0193, Japan \\
$^{2}$ Princeton Plasma Physics Laboratory, Princeton University, Princeton, NJ 08543, USA \\
$^{3}$ The University of Tokyo, Kashiwa, Chiba 277-8561, Japan
}

\date{\today}

 % 1857 / 1920 characters [arxiv limit]
\begin{abstract}\vspace{-0.35cm}
Integrated codes simulating interactions between Alfv\'{e}n waves and fast ions in tokamak plasmas make use of computationally inexpensive perturbative models for the relatively slow processes of instability growth, saturation, chirping and bursting, and associated fast ion transport. The faster processes by which an Alfv\'{e}n mode's spatiotemporal structure forms are assumed to have been completed within the mode's oscillation period, $\tau_0 \equiv 2\pi/\omega_0$. This separation of time scales underlies the computational efficiency of perturbative models, where the Alfv\'{e}n mode's time-dependence is reduced to that of a scalar signal $s(t) = A(t)\sin(-\omega_0 t - \phi(t))$ with variable amplitude $A(t)$ and phase $\phi(t)$. For this, accurate input data in the form of a mode's spatial structure $\delta\Phi({\bm x})$, damping rate $\gamma_{\rm d}$, and initial frequency $\omega_0$ are required, but this is straightforward only for discrete eigenmodes. For modes residing in dense or continuous spectra, $\delta\Phi({\bm x})$ and $\gamma_{\rm d}$ (if necessary, in frequency-dependent form) could be estimated from the form of the continua and fast ion orbits, but it is difficult to guess the correct seed frequency $\omega_0$. Here, we report results of numerical experiments demonstrating that it is possible to find $\omega_0$ by taking advantage of a prompt frequency shift that occurs during the first few $100$ time steps of a simulation. Restarts with the shifted frequency iteratively converge to a value of $\omega_0$ that seems to maximize the resonant drive, suggesting an auto-optimization process. The need for iteration is attributed to the fact that the terms required for rapid frequency adjustments (such as ${\rm d}^2\phi/{\rm d}t^2$) were truncated when deriving the perturbative model. Meanwhile, the fact that partial auto-optimization is possible at all is attributed to the fact that series truncation alone (without low-pass filter) does not strictly enforce slowness. Remnants of and cross-talk with faster dynamics still occur in numerical implementations. This entails potential for both uncertainty and utility.
\end{abstract}

%\pacs{52.55.Fa, 52.55.Pi, 52.35.Bj, 52.35.Mw, 52.65.Ww}

% 52.30.Cv 	Magnetohydrodynamics
% 52.35.Mw 	Nonlinear phenomena: waves, wave propagation, and other interactions (including parametric effects, mode coupling, ponderomotive effects, etc.)
% 52.55.Fa 	Tokamaks, spherical tokamaks
% 52.65.Kj 	Magnetohydrodynamic and fluid equation (Plasma simulation)
% 52.55.Pi 	Fusion products effects (e.g., alpha-particles, etc.), fast particle effects
% 52.65.Ww 	Hybrid methods (Plasma simulation)
% 52.35.-g 	Waves, oscillations, and instabilities in plasmas and intense beams (see also 94.20.wf Plasma waves and instabilities in physics of the ionosphere; 94.30.cq MHD waves, plasma waves, and instabilities in physics of the magnetosphere; 96.50.Tf MHD waves, plasma waves, turbulence in interplanetary physics)
% 52.35.Bj 	Magnetohydrodynamic waves (e.g., Alfven waves)
% 52.35.Dm 	Sound waves
% 52.35.Sb 	Solitons; BGK modes

\maketitle

\thispagestyle{empty}
\everypar{\looseness=-1} % squeeze space

%\tableofcontents

%==============================================================================
\section{Introduction}%\vspace{-0.1cm}
\label{sec:intro}

Resonant interactions with shear Alfv\'{e}n waves play a key role in the regulation of fast ion confinement in a tokamak. The underlying dynamic processes span a wide range of time scales: from rapid mode structure formation (microseconds), via growth and saturation, to chirping and bursting of the modes and transport of the particles in the presence of sources and collisions (milliseconds). These processes can be captured by comprehensive simulation models, but only in certain cases (e.g., few long-wavelength modes) and at a substantial computational cost (days to months on modern supercomputers). Examples can be found in Refs.~\cite{Bierwage17a, Bierwage18, WangJ26}.

Integrated simulation workflows consisting of a chain of specialized codes based on reduced models are an attractive alternative. A high degree of reduction is achieved by models that use information about Alfv\'{e}n eigenmodes (mode structures, frequencies, damping rates) to relax either the fast ion distribution function or its radial density profile via quasilinear diffusion around resonances \cite{Gorelenkov24} or based on the critical gradient paradigm \cite{Bass19}. Here, we are interested in modeling more complex behavior, such as amplitude bursts and dynamic changes in the mode frequency (chirping), which are controlled by resonant wave-particle interactions involving coherent vortical or convective phase space structures \cite{Todo19, HeidbrinkWhite20, WangX22}. Existing workflows with that capability typically consist of an equilibrium solver, a linear eigensolver, and an orbit-following code using Monte-Carlo sampling together with a perturbative model for the resonant dynamics. Examples are the {\tt CASTOR-HAGIS} \cite{Pinches98}, {\tt LIGKA-HAGIS} \cite{Schneller13, Popa23} and {\tt NOVA-TRANSP-ORBIT} \cite{Podesta22} workflows. Currently, work is underway \cite{OyolaTMEP26} to implement such a perturbative mode evolution model also in the widely-used orbit-following Monte-Carlo code {\tt ASCOT5} \cite{Varje19}.

\begin{figure}
	[tbp!]%\vspace{-0.2cm}
	\centering
	\includegraphics[width=0.48\textwidth]{\figures/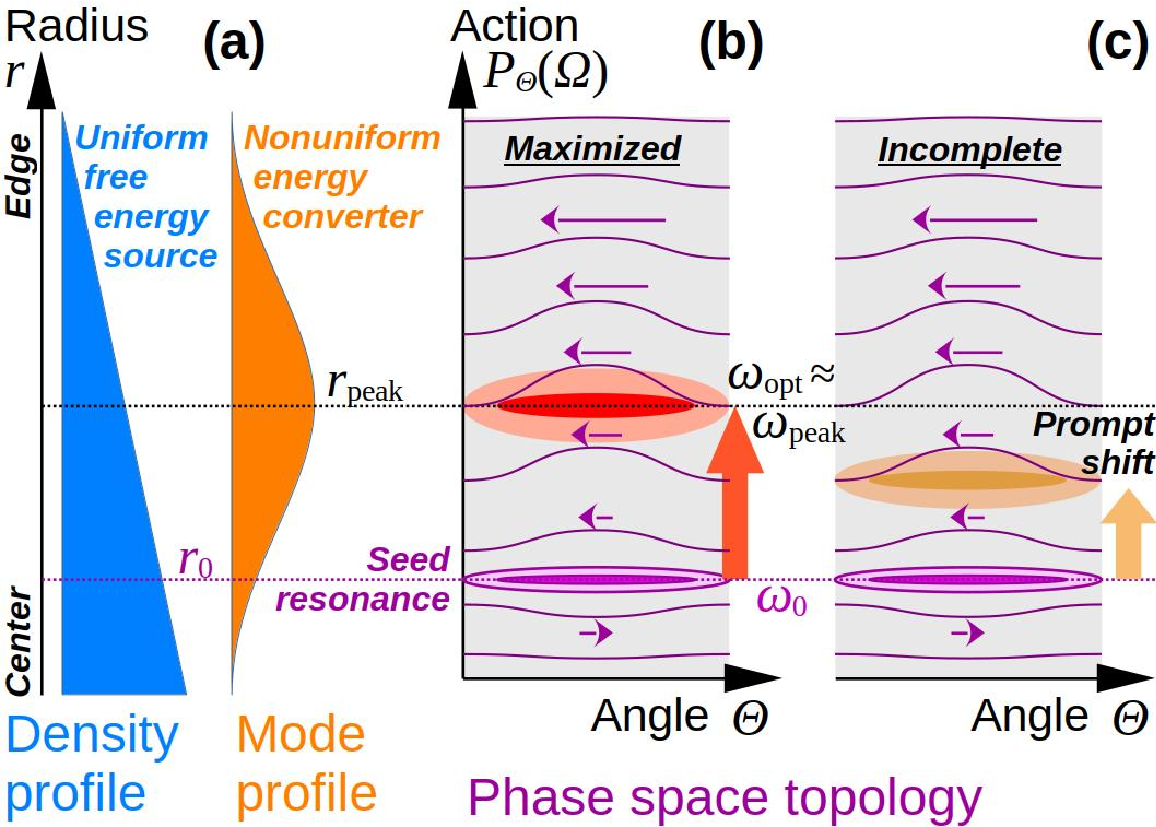}\vspace{-0.2cm}
	\caption{Schematic illustration of (a) our simplified simulation setup for the systematic study of auto-optimized resonant energy conversion, which is expected to be (b) maximized in non-perturbative models, but is (c) incomplete in our perturbative model. This illustration does not incorporate the effect of magnetic drifts, which can cause the optimal resonance to be located off-peak in actual simulations in realistic geometry.}\vspace{-0.4cm}
	\label{fig:01_schematic-cont-res}%
\end{figure}

The existing workflows appear well-suited for ideal MHD Alfv\'{e}n eigenmodes with relatively low amplitudes (i.e., weak MHD nonlinearities), which have well-defined mode structures and frequencies. More problematic are modes residing in dense or continuous spectra, such as ``kinetic'' drift-Alfv\'{e}n modes \cite{Hasegawa75, Cheng85, Mett92} and energetic particle modes (EPM) \cite{Chen94}. In such cases, the process of establishing a mode's spatiotemporal structure cannot be decoupled from the resonant drive \cite{Zonca96a} and, except in special limits \cite{Zonca00}, generally requires an initial value solver. Moreover, when Alfv\'{e}n mode amplitudes become large, they can modify the equilibrium \cite{Chen12}. The development of advanced integrated simulation models incorporating such effects with a high degree of self-consistency is currently an active area of research \cite{Lauber24, Meng24}.

In this paper, we explore ways to enhance the utility of the perturbative model used in codes such as {\tt HAGIS} \cite{Pinches98} and {\tt ORBIT} \cite{ChenY97, ChenY99, White20}, where modes are assumed to retain their spatial structure $\delta\Phi({\bm x})$, while evolving their amplitude $A(t)$ and phase $\phi(t)$. This allows to simulate nonlinear frequency chirping, $\delta\omega(t) = \dot{\phi} \equiv {\rm d}\phi/{\rm d}t$ \cite{Pinches04}. Since this model has no information about the plasma's dielectric response, it relies on accurate inputs in terms of the mode structure and its frequency. As noted above, this information can be provided by linear eigensolvers like {\tt CASTOR}, {\tt LIGKA} or {\tt NOVA}. However, the subsequent nonlinear chirping dynamics are constrained only by the radial width of the prescribed mode and the form of the fast ion orbits, which determine the structure of the resonances. In principle, the missing information about the plasma response could be encoded in a spatially and frequency-dependent damping rate $\gamma_{\rm d}$. Changes in the mode's spatial structure $\delta\Phi({\bm x})$ could also be modeled based on information about the continuous spectra and magnetic drifts. To our knowledge, such models do not exist, so we currently operate with fixed $\gamma_{\rm d}$ and $\delta\Phi({\bm x})$.

On the one hand, this limits the present model's validity to cases with Alfv\'{e}n eigenmodes that perform no or only little chirping. In such cases, one may as well exclude the chirping altogether and evolve only the amplitude $A(t)$ while fixing the frequency ($\dot{\phi} = {\rm const}$.), as is the case in quasilinear relaxation models \cite{Gorelenkov24,Bass19}.

On the other hand, the lack of constraints on frequency shifts $\delta\omega(t)$ could be seen as making such a perturbative model suitable for simulating the interaction between fast ions and an Alfv\'{e}n mode residing in a dense spectrum or continuum, at least in some cases. A concrete example that comes to mind are long-wavelength EPMs that were seen in beam-driven JT-60U tokamak plasmas \cite{Bierwage17a, Bierwage18}. The pronounced beating (via linear interference) that was observed in both experimental measurements and simulations \cite{Bierwage17a} indicated that modes with approximately the same spatial structure were excited across a relatively wide range of frequencies; in that case $(50\pm 10)\,{\rm kHz}$. These conditions appear suitable for being simulated with a perturbative model, and we have previously done so in Ref.~\cite{Bierwage21}, where we studied the effect of beating on wave-particle trapping using the {\tt ORBIT} code.

In the model we used, modes with similar spatial structure were represented by a single wavefunction $\delta\Phi({\bm x})$ that is capable of oscillating at multiple frequencies simultaneously (like the membrane of a loudspeaker) in response to dynamic forcing from the fast ion population (acting like an orchestra). Like any initial-value simulation, the dynamics have to be kick-started with a symmetry-breaking perturbation. A simulation of actual MHD modes is best initialized with random ``noise'' in order to ensure that the fastest growing mode --- the one that maximizes the energy conversion rate --- will be the first to emerge from the incoherent fluctuations. In contrast, our current perturbative model lacks the physics that govern the spontaneous auto-organization of the wavefield self-consistently, so one has to precommit to a set of modes; in our case, only one mode. 

Our expectation had been that, given a spatial structure $\delta\Phi({\bm x})$ with small initial amplitude $A_0$ and a seed frequency $\omega_0$, we would effectively kick-start a narrow set of Van-Kampen modes \cite{VanKampen55} around a frequency $\Omega$ satisfying the approximate resonance condition $|\omega_0 - \Omega| \lesssim |\gamma|$ that may be broadened or shifted by an amount comparable to the net growth rate $\gamma$, which can be seen as a collisionless linear resonance width.

To our surprise, whenever there was a spatial separation between the prescribed mode's peak $r_{\rm peak}$ and the seed resonance radius $r_0(\omega_0)$ as in Fig.~\ref{fig:01_schematic-cont-res}(a), the simulations in Ref.~\cite{Bierwage21} began with a prompt frequency shift $\omega_0 \rightarrow \omega_0 + \delta\omega$ that depended mainly on the distance $r_{\rm peak}-r_0$, while the dependence on the rates of drive and damping (and thus $\gamma$) was weak. This observation was tentatively interpreted as an ``auto-optimization of resonant drive'', which was also the title of that paper's Appendix~D.4, where this phenomenon was described. It is this presumed automatic correction of an inaccurate guess $\omega_0$ of the optimal resonant frequency $\omega_{\rm opt}$ that we shall focus on in the present paper, whose purpose is to
\begin{enumerate}
	\item[1.] clarify the properties of the prompt frequency shift within the scope of the present perturbative model,
\end{enumerate}
\noindent and, based on that understanding and evidence,

\begin{enumerate}
	\item[2.] demonstrate that this prompt shift can be used to improve our initial guess for the frequency $\omega_0$ of the resonance that will act as the seed for subsequent nonlinear chirping in a perturbative model.
\end{enumerate}

The auto-optimization paradigm and the general problem setting for this study are described in the following Section~\ref{sec:auto}. The reduced model and the underlying time-scale separation problem are described in Section~\ref{sec:model}. The concrete simulation setup used in this work is described in Section~\ref{sec:setup}. The results of several parameter scans that we found insightful are presented and discussed in Section~\ref{sec:results}, followed by a summary and conclusions in Section~\ref{sec:summary}. The Appendix contains supplementary information concerning resonance conditions.
 
%==============================================================================
\section{Problem setting}\vspace{-0.3cm}
\label{sec:auto}

The concrete question that motivated the present study is the following: {\it Does the prompt frequency shift in our perturbative simulations \cite{Bierwage21} really maximize the energy conversion rate as expected for the dominant linear instability?} Based on the numerical experiments that will be reported below, the answer is: {\it Not completely, because restarting the simulation with the shifted frequency results in another, albeit smaller shift.} Before proceeding to the details, let us outline the situation in simplified terms using the schematic illustration in Fig.~\ref{fig:01_schematic-cont-res}.

The blue triangle in Fig.~\ref{fig:01_schematic-cont-res}(a) represents the fast ion density profile with constant gradient along the minor radial coordinate $r$ of the torus. This gradient serves as a uniform source of free energy. The background plasma, serving as the wave-carrying medium, is assumed to support a mode that peaks at a certain radius $r_{\rm peak}$. This mode acts as a spatially nonuniform energy converter whose radial profile is drawn as a bell-shaped orange area in Fig.~\ref{fig:01_schematic-cont-res}(a). While the mode structure is fixed, its overall amplitude $A(t)$ and phase $\phi(t)$ can vary under the influence of wave-particle interactions, giving the signal
\begin{equation}
	s(t) = A(t) \sin(-\omega_0 t - \phi(t)),
	\label{eq:intro_signal}
\end{equation}

\noindent where $\omega_0$ is the initial frequency.

Fig.~\ref{fig:01_schematic-cont-res}(b) shows a two-dimensional projection of the phase space populated by passing particles that can interact with the mode. In the absence of perturbations ($A=0$), each particle moves freely along the horizontal axis of the angle $\Theta$ with constant angular momentum $P_\Theta$. Here, particles at different radii $r$ are assumed to have different angular momenta $P_\Theta(r)$, which translates into different characteristic transit frequencies $\Omega(r)$ along the vertical axis. After the mode is introduced with a small initial amplitude $0 < A_0 \equiv A(t=0) \ll A(t_{\rm sat})$ (saturation level), particles with $\Omega(r_0) = \omega_0$ become resonant. The phase space in Fig.~\ref{fig:01_schematic-cont-res}(b) now consists of topologically distinct domains of untrapped (magenta arrows) and trapped particle orbits (magenta ellipse).

Our setup in Fig.~\ref{fig:01_schematic-cont-res} is such that the seed resonance lies far from the mode's peak. This means that the wave-particle energy transfer rate will be maximized when the mode changes its frequency from $\omega_0$ to the optimal value $\omega_{\rm opt} \approx \omega_{\rm peak} \equiv \omega_0 + \delta\omega_0$ satisfying $\omega_{\rm peak} = \Omega(r_{\rm peak})$, as indicated by the red arrow in Fig.~\ref{fig:01_schematic-cont-res}(b). The shifted frequency is then maintained during the entire exponential growth phase; that is, after the prompt relaxation on the short time scale $\pi/\omega_0$, and before nonlinear saturation at $t_{\rm sat}$, independently of the mode amplitude $A(t)$:
\begin{equation}
	\delta\omega_0 \equiv \dot{\phi}(\pi/\omega_0 \lesssim t \ll t_{\rm sat}) \approx {\rm const}.
\end{equation}

If the system would be capable of realizing complete auto-optimization as in Fig.~\ref{fig:01_schematic-cont-res}(b), then the result $\omega_0 + \delta\omega_0$ should be independent of the initial frequency $\omega_0$. However, in our perturbative simulations, the prompt frequency shift is found to fall short of complete optimization as illustrated in Fig.~\ref{fig:01_schematic-cont-res}(c). This will be verified below by restarting the simulation with a new initial frequency $\omega_0' = \omega_0 + \delta\omega_0$. The occurrence of yet another (smaller) prompt shift implies incomplete auto-optimization.

While the optimal choice for $\omega_0$ is self-evident in the simple setup shown in Fig.~\ref{fig:01_schematic-cont-res}, the situation becomes more interesting when magnetic drifts are sufficiently large to shift the optimal resonance off-peak and when the fast ion gradients are nonuniform and steepest at an off-peak location. Such a situation may occur, for instance, in the case of kink modes whose resonant interaction with fast ions can give rise to ``fishbone'' bursts \cite{McGuire83,Chen84}, including recently discovered double-peaked variants \cite{Lee23,Lee26}.

In order to facilitate a systematic and transparent analysis with clear results, our simulation setup described in the following Section~\ref{sec:model} employs some of the simplifications and exaggerations of Fig.~\ref{fig:01_schematic-cont-res}. At the same time, we will use a realistic magnetic geometry and include the effect of magnetic drifts.

%==============================================================================
\section{Model \& scale separation problem}
\label{sec:model}

We consider a tokamak plasma with an ambient magnetic field ${\bm B} = \nablab\zeta\times\nablab\Psi_{\rm P} + g(\Psi_{\rm P})\nablab\zeta$ that is symmetric in the toroidal angle $\zeta$ and has singly nested toroidal surfaces of constant magnetic flux $2\pi\Psi_{\rm P}$. The total magnetic field ${\bm B}_{\rm tot} = \nablab\times{\bm A}_{\rm tot} = {\bm B} + \delta{\bm B}$ contains a fluctuating component that satisfies Faraday's law $\partial_t\delta{\bm B} = -\nablab\times\delta{\bm E}$ with electric field $\delta{\bm E} = -\nablab\delta\Phi - \partial_t\delta{\bm A}$, electrostatic potential $\delta\Phi({\bm x},t)$ and vector potential $\delta{\bm A}({\bm x},t)$. As a starting point for the fluctuations, we assume an Alfv\'{e}nic wave field that satisfies the ideal MHD constraint $\delta E_\parallel \approx \hat{\bm b}\cdot\delta{\bm E} = -\partial_t\delta A_\parallel - \nabla_\parallel\delta\Phi \approx 0$ with $\hat{\bm b} \equiv {\bm B}/B$ and $B \equiv |{\bm B}|$, has a small magnitude $\delta B/B \ll 10^{-2}$, and a negligible magnetoacoustic component $\delta B_\parallel \equiv \hat{\bm b}\cdot\delta{\bm B} \ll \delta B$, so that we may write $\delta{\bm B} \approx \nablab\times\alpha{\bm B}$ with $\alpha \equiv \delta A_\parallel/B$ and $\nabla_\parallel\delta\Phi \approx -B\partial_t\alpha$. In other words, we consider incompressible shear Alfv\'{e}n waves in a nonuniform plasma. In our case, this assumption is justified for modes with frequencies $> 30\,{\rm kHz}$ (above the kinetic thermal ion gap in Fig.~7 of Ref.~\cite{Bierwage17a}), but is expected to become invalid for fishbone-type EPMs and other low-frequency modes, which have a significant compressional component \cite{Du24}.

In order to study wave-particle resonances in realistic tokamak geometry in an inexpensive, systematic and transparent way, we employ a perturbative reduced model, where Hamiltonian guiding center equations are coupled to an Alfv\'{e}n-mode-like wave field, whose electrostatic potential is assumed to have the following form:
\begin{equation}
	\delta\Phi = \underbrace{A_0\delta\tilde{\Phi}(r) e^{in\zeta - im\vartheta - i\omega_0 t}}\limits_{\text{prescribed seed wave}} \underbrace{\hat{A}(t) e^{-i\phi(t)}}\limits_{\text{dynamic part}} - {\rm c.c.},
	\label{eq:mode}
\end{equation}

\noindent where $r(\Psi_{\rm P})$ is a minor radial coordinate, $\vartheta$ and $\zeta$ are the poloidal and toroidal angles in Boozer coordinates, and $\hat{A} \equiv A/A_0$. The seed wave has a user-defined initial amplitude $A_0$ and frequency $\omega_0$. The latter can be absorbed into the toroidal angle by going into a rotating frame of reference ($\zeta' = \zeta - \omega_0 t/n$), where wave-particle interactions can be studied conveniently. The model was first used in Refs.~\cite{ChenY97,Pinches98,ChenY99}. The numerical implementation in the code {\tt ORBIT} used here was described in Refs.~\cite{White20, Bierwage21, Bierwage22c}.\footnote{Governed by the Vlasov equation and Hamiltonian guiding center equations --- Eqs.~(4), (18), (19) of Ref.~\protect\cite{Bierwage21}, and details in Appendix A.4 of Ref.~\protect\cite{Bierwage22c} --- the guiding center distribution $f_{\rm gc}(t)$ evolves under the influence of the fluctuating electromagnetic field (\protect\ref{eq:mode}). The evolution of the mode amplitude $A(t)$ and phase $\phi(t)$ under the influence of $f_{\rm gc}(t)$ is, in turn, governed by the perturbative model appearing in Eqs.~(2) and (15) of Ref.~\protect\cite{Bierwage21}.}

The perturbative character of this model lies in the fact that the mode's dynamics are simplified as follows:
\begin{enumerate}
	\item[(I)]  The radial mode structure $\delta\tilde{\Phi}(r)$ for each set of poloidal and toroidal Fourier mode numbers $(m,n)$ as well as the mode number spectrum are prescribed input parameters and do not evolve in time.
	\item[(II)]  The evolution of the frequency $\omega(t) = \omega_0 + \dot{\phi}(t)$ is not subject to any constraints from the bulk plasma response function. $\omega_0$ is an input parameter, and $\phi(t)$ is determined by interactions with fast ions.
	\item[(III)]  The mode is subject to $\omega$- and $t$-independent, spatially uniform damping $0 \leq \gamma_{\rm d} = {\rm const}$.
	\item[(IV)]  The amplitude modulations $A(t)$ and dynamic phase shifts $\phi(t)$ are assumed to vary slowly in the sense that ${\rm d}\ln|A|/{\rm d}t \ll \omega_0$ and ${\rm d}\phi/{\rm d}t \ll \omega_0$ \cite{Pinches98,ChenY99}.
\end{enumerate}

\noindent The user is responsible for providing sensible inputs for (I)--(III). Concerning (IV), it turns out that the $t$-scale-separation assumption is only imperfectly enforced in actual numerical implementations of this model. This imperfection entails potential for both uncertainty and utility. In our case, it will provide constraints and guidance for our choice of $\omega_0$ in (II). Therefore, let us discuss the imperfect $t$-scale-separation problem in more detail.

Starting from the momentum balance equation of the kinetic-MHD hybrid model in current-coupling form \cite{Park92}, Y.Chen \& R.B.White \cite{ChenY99} use the above assumptions to reduce the original second-order partial differential equation to a pair of first-order ordinary differential equations for $A(t)$ and $\phi(t)$:\vspace{-0.1cm}
\begin{subequations}
	\begin{align}
		\frac{{\rm d}A}{{\rm d}t} =&\, -\frac{C_A^2}{A\omega_0^2} \left<\int{\rm d}{\bm x}\, {\bm J}_{\rm f}\cdot\delta{\bm E}_\perp\right>_{\tau_0} - \gamma_{\rm d} A, \\
		A\frac{{\rm d}\phi}{{\rm d}t} =&\, -\frac{C_A^2}{A\omega_0^3} \left<\int{\rm d}{\bm x}\, {\bm J}_{\rm f}\cdot\partial_t\delta{\bm E}_\perp\right>_{\tau_0},
	\end{align}
	\label{eq:pert_A_phi}
\end{subequations}\vspace{-0.2cm}

\noindent where ${\bm J}_{\rm f}$ is the fast ion current density, $\delta{\bm E}_\perp$ the mode's fluctuating electric field component transverse to the ambient magnetic field ${\bm B}$. The amplitude $A$ is related to the ``radial'' displacement in the space of toroidal flux $\Psi$ as $A\hat{\xi}^\Psi = {\bm \xi}\cdot\nablab\Psi/B_0$, where $B_0$ is the on-axis magnetic field strength and ${\bm \xi}$ the spatial displacement vector. The constant $C_A$ in Eq.~(\ref{eq:pert_A_phi}) is such that $\hat{\xi}^\Psi$ has a peak of unity as in Eq.~(\protect\ref{eq:model_xi}) below. Eq.~(\ref{eq:pert_A_phi}) is a compact form of Eqs.~(6.80) and (6.81) in the textbook by R.B.White \cite{WhiteTokBook3}.

What we would like to draw the reader's attention to are the integrals $\left<...\right>_{\tau_0}$ over the fast time scale $\tau_0 \equiv 2\pi/\omega_0$ of the wave oscillation period. When suitably weighted, such averages are equivalent to a low-pass filter that permits only dynamics on time scales longer than $\tau_0$ to influence the evolution of $A$ and $\phi$. If one would actually perform those integrals in the simulation, the amplitude deaths and phase jumps that constitute linear beating (studied in Ref.~\cite{Bierwage21}) would not happen, or perhaps only in smoothed, attenuated form. The same counts for the prompt frequency shift $\delta\omega_0$ studied here.

However, {\tt ORBIT} does not actually make use of those time-averaging integrals. Instead, like the {\tt HAGIS} code \cite{Pinches98}, {\tt ORBIT} relies wholly on analytical truncation, where higher-order derivatives and products like $\ddot{\phi}$, $\ddot{A}$, $\dot{\phi}\dot{A}$ are excluded. This is usually safe in mathematical physics studies, but it can be a source of uncertainty in numerical implementations thereof, as we will discuss shortly. The resulting equations in {\tt ORBIT} have the form \cite{White20, Bierwage21}
\begin{subequations}
	\begin{align}
		\frac{{\rm d}A}{{\rm d}t} =&\, -\frac{v_{\rm A}^2}{\omega_0 D} \sum_k w_k \left[\left(\varrho_\parallel B^2 \alpha - \delta\Phi\right)\cos{\Theta}_{mn} \right]_k \\
		&\,- \gamma_{\rm d} A, \nonumber \\
		\frac{{\rm d}\phi}{{\rm d}t} =&\, -\frac{v_{\rm A}^2}{A\omega_0 D} \sum_k w_k \left[\left(\varrho_\parallel B^2 \alpha - \delta\Phi\right)\sin{\Theta}_{mn}\right]_k,
	\end{align}
	\label{eq:pert_A_phi_unfilt}
\end{subequations}\vspace{-0.2cm}

\noindent where $D$ is a normalization factor, and where we have discretized the phase space integrals and represented them as sums over simulation particles $k$ with $\delta f$-weight $w_k$, charge-mass ratio $\tfrac{Ze}{M}$, parallel guiding center velocity $u$, parallel gyroradius $\varrho_\parallel \equiv \frac{M u}{ZeB}$, magnetic moment $\mu$, and wave-frame phase $\Theta_{mn} \equiv n\zeta - m\vartheta - \omega_0 t - \phi(t)$.

Unlike a (suitably weighted) time-average as $\left<...\right>_{\tau_0}$ in Eq.~(\ref{eq:pert_A_phi}), the scale separation via ordering-and-truncation that underlies Eq.~(\ref{eq:pert_A_phi_unfilt}) does not perform as a proper low-pass filter. Presumably, this is true for any model derived by similar means, and it can be a source of concern as it is not clear how the slow dynamics of interest are affected by cross-talk with distorted remnants of faster processes.

However, this systematic caveat may be tolerable if the physical nature of the system at hand guarantees that
\begin{enumerate}
	\item[(IV-1)] fast initial transients have no long-term effect, and
	\item[(IV-2)] the system does not generate any further fast dynamics once a proper solution has been established.
\end{enumerate}

\noindent Fast initial transients arise when the symmetry-breaking initial perturbation does not constitute an exact solution of Eq.~(\ref{eq:pert_A_phi_unfilt}), as is always the case in our simulations since we perturb only the electromagnetic fields, not the particle distribution. The initial transients correct the inconsistencies in all available degrees of freedom, which in the (amplitude-independent) linear regime comprise the guiding center distribution $f_{\rm gc}(\psi_{\rm P},\vartheta,\zeta,\varrho_\parallel,t|\mu)$ and the phase $\phi(t)$. Thus, if we provide a ``wrong'' initial frequency $\omega_0$, which does not maximize the energy conversion rate, the system can establish a phase drift $\delta\omega_0 \equiv \dot{\phi}$ that corrects this inconsistency. However, we will find that this prompt frequency shift $\delta\omega_0$ falls short of fully optimizing the resonant drive. We attribute this to the absence of the above-mentioned ``fast terms'', like $\ddot{\phi}$.

We will also find that our transients violate condition (IV-1): The ensuing prompt frequency shift $\delta\omega_0$ is maintained throughout the resonant instability's exponential growth, and even the spectral band occupied by the subsequent nonlinear dynamics is going to be visibly shifted. In other words, the system permanently retains memory of the initial condition $\omega_0$.

As for condition (IV-2), we know for a fact that there are two phenomena that violate it at least formally: particle-in-cell (PIC) noise and linear beating. Effects of undesirable signal-noise correlations can be kept under control by implementing a so-called quiet start and by using a sufficiently large number of simulation particles (based on numerical convergence tests). The role of linear beating is elucidated in the following paragraphs from multiple perspectives.

First, linear beating formally violates condition (IV-2) if one directly solves Eqs.~(\ref{eq:pert_A_phi_unfilt}) for $A(t)$ and $\phi(t)$. One way to see this is by noting that these variables are equivalent to polar coordinates in the complex plane,
\begin{equation}
	A(t) e^{-i\phi} = X(t) - iY(t),
\end{equation}

\noindent and are thus ill-behaved in near the origin, $A \rightarrow 0$, whose neighborhood the system traverses between beats.

Evidently, this problem can be avoided through the use of the auxiliary variables $X(t) = A(t)\cos\phi$ and $Y(t) = A(t)\sin\phi$ \cite{Pinches98}. These Cartesian coordinates vary smoothly throughout the complex plane, even during simple linear beats. We may thus assume that condition (IV-2) is not truly violated by linear beating, at least as long as it involves only two or perhaps a few superimposed waves. This situation is characteristic for the early nonlinear phase of a collisionless resonant instability that involves hole-clump pair formation\footnote{For instance, see Fig.~17 of Ref.~\cite{Bierwage21}, and $t < 1\,{\rm ms}$ in Fig.~\ref{fig:04_scanr}(A,B).}
(to which we shall return to shortly). In this regime, we may assume that even if we use Eq.~(\ref{eq:pert_A_phi_unfilt}) for $A(t)$ and $\phi(t)$, the issues, if any, will be primarily numerical, not physical.

Next, let us consider the advanced nonlinear regime, where the beats can become chaotic, so that one may suspect that even $X(t)$ and $Y(t)$ may have episodes of rapid variation violating condition (IV-2). Here, we find it useful to invoke the physical picture of linear beating as described in Ref.~\cite{Bierwage21} in terms of ``density waves''  (holes \& clumps) in guiding center phase space. By summation over the simulation particles representing these phase space structures, Eq.~(\ref{eq:pert_A_phi_unfilt}) is capable of reproducing perfect linear beats, with abrupt, step-function-like changes in the growth rate $\gamma(t) = {\rm d}\ln A/{\rm d}t$ and phase jumps $\Delta\phi = \pm\pi$ \cite{White19,White20,Bierwage21}. However, these rapid variations of $A(t)$ and $\phi(t)$ during linear beating appear only in the virtual carrier wave (which does not exist in the sense that there is no associated spectral peak). Meanwhile, the interfering physical waves --- here hole \& clump vortices in phase space, and the associated spectral peaks in the field response $\tilde{\Phi}(t) \propto s(t)$ in Eq.~(\ref{eq:intro_signal}) --- satisfy the slowness condition (IV-2). This has two reasons:
\begin{itemize}
	\item  All phase space structures are composed of marker particles whose $\delta f$-weights and trajectories vary smoothly while obeying Liouville's theorem (incompressibility of phase space).
	\item  Coherent phase space structures of significant size and, thus, with significant influence arise only near resonances, $\Omega \approx \omega_0$, which means that they are guaranteed to evolve slowly relative to $\tau_0 \equiv 2\pi/\omega_0$ in the wave's frame of reference.
\end{itemize}

\noindent The first bullet item follows from the fact that Eq.~(\protect\ref{eq:pert_A_phi_unfilt}) can be derived from a long-time-scale Lagrangian formulation of the wave field as described by S.D.~Pinches {\it et al}.\ in Ref.~\protect\cite{Pinches98}; namely, the ``wave Lagrangian $\L_{\rm w}$'' in Eq.~(17) of that paper. Short-time-scale dynamics are, in principle, fully retained in the guiding phase space distribution (``fast particle Lagrangian $\L_{\rm fp}$'') and wave-particle coupling (``interaction Lagrangian $\L_{\rm int}$''), but based on the second bullet item, we may expect that only dynamics satisfying the slowness condition (IV-2) will occur. The summation on the right-hand side of Eq.~(\ref{eq:pert_A_phi_unfilt}) can still turn slowly evolving waves into fast beats by linear superposition, but as we noted earlier, those beats should be inconsequential for time scale separation considerations as long as the underlying superimposed waves evolve slowly.

In summary, although it is difficult to rule out spurious effects with absolute certainty, there are reasons to believe that condition (IV-2) for the slowness of self-consistent dynamics will be satisfied, or that violations thereof may at least be tolerable for our purposes.\footnote{A better understanding of remnants of the imperfect $t$-scale separation in Eq.~(\protect\ref{eq:pert_A_phi_unfilt}) may also help to throw more light on
puzzling features that were seen in the long-time evolution of the guiding center distribution in Figs.~29, C1, C2 of Ref.~\protect\cite{Bierwage21}, as well as observations of ``ghost chirps'' described in Section~C.4 of \protect\cite{Bierwage21}.}
The main source of concern is then the violation of condition (IV-1); namely, the long-term effects of fast initial transients. The prompt frequency shift $\delta\omega_0$ studied in the present paper falls in this category. However, based on our interpretation of $\delta\omega_0$ as having a physical origin --- namely, auto-optimization of resonant drive --- we claim that this apparent ``bug'' may turn out to be a useful feature.

%==============================================================================
\section{Simulation setup}
\label{sec:setup}

\begin{figure}
	[tbp!]
	\centering
	\includegraphics[width=0.48\textwidth]{\figures/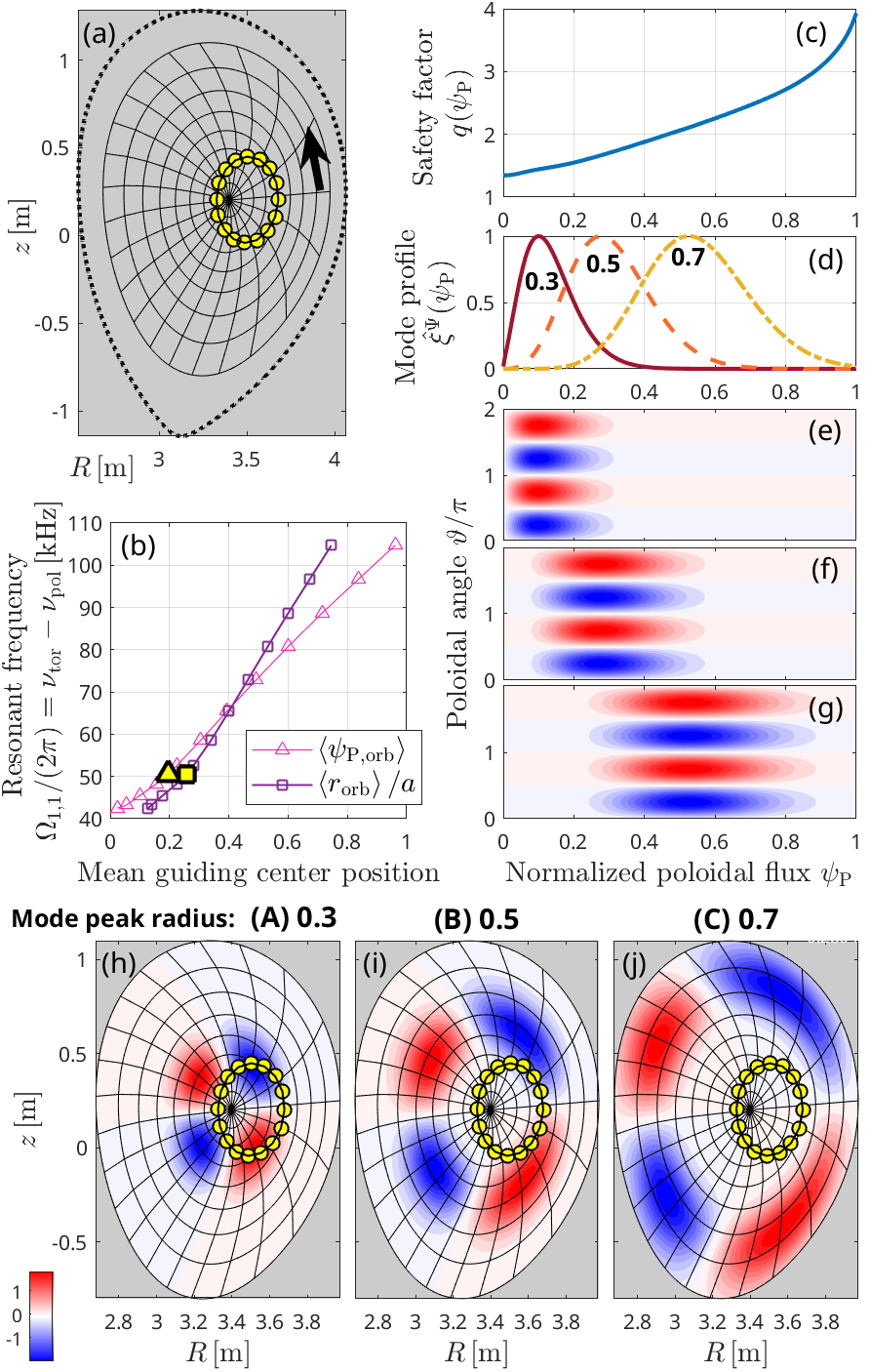}%\vspace{-0.25cm}
	\caption{Simulation setup based on a JT-60U tokamak plasma. (a): Boozer coordinate mesh (solid black), plasma boundary (dotted black), reference resonance (black-yellow circles), and direction of co-passing guiding center motion (black arrow). Plasma parameters: current $I_{\rm p} = 0.57\,{\rm MA}$, on-axis field $B_0 = 1.166\,{\rm T}$, magnetic axis location $(R_0, z_0) = (3.40\,{\rm m},0.20\,{\rm m})$, minor radius $a = 0.65\,{\rm m}$, on-axis Alfv\'{e}n frequency $\omega_{\rm A0} \equiv v_{\rm A0}/R_0 = 2\pi\times 203.89\,{\rm kHz}$. Resonant orbit example: kinetic energy $K = M v^2/2 = 400\,{\rm keV} = 0.4939\times Mv_0^2$ (with reference energy $Mv_0^2/2 = 404.9\,{\rm keV}$), pitch coordinate $\mu B_0 = 167.5\,{\rm keV}$ ($u/v = 0.7830 = \sin(0.2863\pi)$), low-field-side launch point $R_{\rm lfs} = R_0 + 0.28\,{\rm m}$ ($r_{\rm lfs}/a = 0.36$) near the outer midplane ($z_{\rm lfs} = z_0$). (b): Radial dependence of the unperturbed resonant frequency $\Omega_{p,n} = n\omega_{\rm tor} - p\omega_{\rm pol}$ for $p/n = 1/1$ and fixed $K$ and $u_{\rm lfs}/v$, plotted as a function of the mean guiding center position in flux space, $\left<\psi_{\rm P,orb}\right>$ (triangles), and volume-averaged minor radius, $\left<r_{\rm orb}\right>/a$ (squares). The innermost sample is a point-like stagnation orbit. The resonance with $\Omega_{1,1} \approx 0.2489\omega_{\rm A0} \approx 2\pi\times 50.6\,{\rm kHz}$ at $\left<\psi_{\rm P,orb}\right> \approx 0.20$ and $\left<r_{\rm orb}\right>/a \approx 0.26$ is highlighted by large yellow-black symbols. (c): Safety factor profile $q(\psi_{\rm P})$. (d): Radial displacement profiles $\hat{\xi}^\Psi(\psi_{\rm P})$ given by Eq.~(\protect\ref{eq:model_xi}), with half-width $w_0/a = 0.15/\sqrt{2}$ and peak radii (A) $\rpeak/a = 0.3$, (B) $0.5$, and (C) $0.7$. (e)--(j): $m/n = 2/1$ mode structures in the $(\psi_{\rm P},\vartheta)$-plane and in the $(R,z)$-plane. The latter is overlaid with the yellow-black resonance contour of the example from panel (a).}%\vspace{-0.2cm}
	\label{fig:02_setup}%
\end{figure}

Our working scenario is a JT-60U tokamak plasma for which comprehensive long-time simulations \cite{Bierwage17a,Bierwage18}  have previously reproduced experimentally observed chirping and beating energetic particle modes (EPM \cite{Chen94}) and Abrupt Large-amplitude Events (ALE \cite{Shinohara01}). The plasma shape and the Boozer coordinate mesh in the poloidal plane are shown in Fig.~\ref{fig:02_setup}(a) and the profile of the safety factor $q$ is plotted in (c). Relevant plasma parameters are summarized in the caption of Fig.~\ref{fig:02_setup}, and the full set of profiles can be found in Fig.~1 of Ref.~\cite{Bierwage22c}.

The original comprehensive simulations utilized the kinetic-MHD hybrid model of Ref.~\cite{Park92} in the current-coupling form, combining visco-resistive full MHD equations with drift-kinetic equations for guiding centers representing (mostly co-passing) energetic deuterons from negative-ion-based neutral beams with energy $400\,{\rm keV}$. The most important modes had long wavelengths with toroidal Fourier mode numbers $n=1,2,3$. Their mode structures and resonances were characterized in Figs.~8, 12 and 18 of Ref.~\cite{Bierwage14}.

Here, we will consider deeply co-passing deuterons with mass $M = 2 M_{\rm p}$ charge $Ze = e$, kinetic energy $335 \,{\rm keV} \lesssim K \equiv M v^2/2 \lesssim 410\,{\rm keV}$ and pitch parameter $\mu B_0 = 167.5\,{\rm keV}$, and simulate their interactions with a mode of the simplified form given by Eq.~(\ref{eq:mode}). Our test cases have only a single poloidal and toroidal Fourier component, $m/n = 2/1$, and a Gaussian profile in the minor radial coordinate $r$:%\vspace{-0.15cm}
\begin{equation}
	\delta\tilde{\Phi}(r) \propto \hat{\xi}^\Psi(r) = \exp\left(-\frac{(r-\rpeak)^2}{2 w_0^2}\right).
	\label{eq:model_xi}
\end{equation}%\vspace{-0.25cm}

\noindent The relation between $\delta\tilde{\Phi}$ and the radial displacement $\hat{\xi}^\Psi$ and magnetic fluctuation $\alpha$ is given in Eq.~(14c) of Ref.~\cite{Bierwage21}. The half-width of $\hat{\xi}^\Psi(r)$ will be fixed at $w_0/a = 0.15/\sqrt{2} \approx 0.11$. For the peak location, we will consider three cases: (A) $\rpeak/a = 0.3$, (B) $0.5$ and (C) $0.7$. The corresponding radial profiles plotted as functions of the normalized poloidal flux $\psi_{\rm P}(r) = \Psi_{\rm P}(r)/\Psi_{\rm P,edge} \in [0,1]$ and global mode structures in the poloidal $(R,z)$ plane are shown in Fig.~\ref{fig:02_setup}(d)--(j).

The default seed frequency is $\nu_0 \equiv \omega_0/(2\pi) \approx 50.6\,{\rm kHz}$, with scans ranging up to $80\,{\rm kHz}$. The guiding center orbit populated by $400\,{\rm keV}$ deuterons whose toroidal and poloidal orbital frequencies $\omega_{\rm tor}$ and $\omega_{\rm pol}$ as defined in Eq.~(\ref{eq:wtrans}) satisfy the resonance condition of the form%\vspace{-0.1cm}
\begin{equation}
	\omega = \Omega_{p,n} \quad \text{with} \quad \Omega_{p,n} \equiv n\omega_{\rm tor} - p\omega_{\rm pol},
\end{equation}%\vspace{-0.5cm}

\noindent where $p=1$, $n=1$ and $\omega= \omega_0$ are shown in Fig.~\ref{fig:02_setup}(a) as yellow-black circles. This drift orbit contour represents the shape of our reference resonance in configuration space $(R,z)$. The magnetic field and plasma current both point out of the plane, so the co-passing guiding centers ${\bm X}_{\rm gc}(t)$ propagate counter-clockwise as indicated by the black arrow. The guiding center velocity $\dot{\bm X}_{\rm gc}(t) \equiv {\rm d}{\bm X}_{\rm gc}(t)/{\rm d}t$ has the parallel component%\vspace{-0.15cm}
\begin{equation}
	u(t) = \sigma\sqrt{\tfrac{2}{M}\left(K(t) - \mu B({\bm X}_{\rm gc}(t))\right)} \equiv \hat{\bm b}\cdot\dot{\bm X}_{\rm gc},
	\label{eq:u}
\end{equation}\vspace{-0.35cm}

\noindent with sign $\sigma \equiv \hat{\bm b}\cdot\dot{\bm X}_{\rm gc}/|\hat{\bm b}\cdot\dot{\bm X}_{\rm gc}|$.

Panels (h)--(j) of Fig.~\ref{fig:02_setup} show that this $50.6\,{\rm kHz}$ reference resonance overlaps with a large part of mode (A), and the low-field-side portion of mode (B), while residing on the far inner outskirts of mode (C). Case (C) is an extreme example that will allow us to demonstrate the effect of interest --- namely, a prompt frequency shift towards the mode's peak --- most clearly.

\begin{figure}
	[tbp]
	\centering
	\includegraphics[width=0.48\textwidth]{\figures/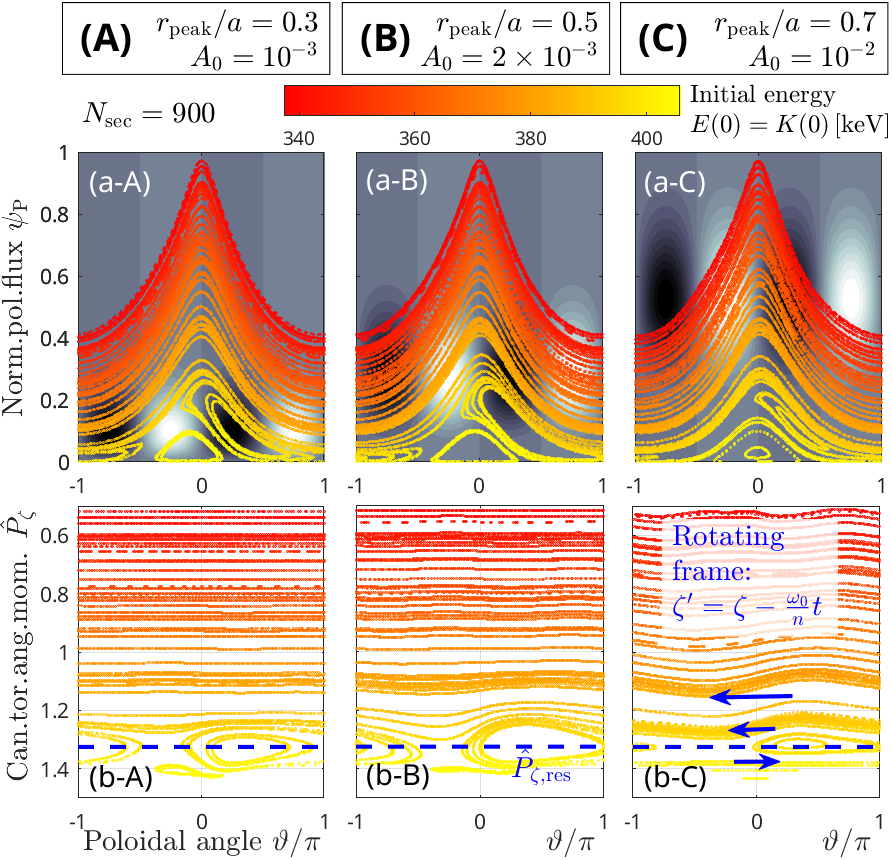}%\vspace{-0.2cm}
	\caption{Poincar\'{e} plots in the toroidally rotating plane $\zeta' = \zeta  - \omega_0 t/n$ with $n=1$ and $\omega_0 \approx 50.6\,{\rm kHz}$, showing the domain covered by simulation particles and the structure of the unshifted ($\dot{\phi}=0$) seed resonance for the three cases of Fig.~\protect\ref{fig:02_setup}, where the mode's peak is located at (A) $\rpeak/a = 0.3$, (B) $0.5$, and (C) $0.7$. (a): Mode structure $\delta\hat{\xi}^\Psi(\psi_{\rm P}) e^{-im\vartheta}$ (gray-scale) in the $(\psi_{\rm P},\vartheta)$-plane, overlaid with Poincar\'{e} plots from passive test particles, colored according to their initial kinetic energy $K(t=0)$ ranging from $335\,{\rm keV}$ near the plasma boundary at the outer edge of the outer midplane, $(\psi_{\rm P},\vartheta) \approx (1,0)$ (red), to nearly $410\,{\rm keV}$ in the plasma center, $(\psi_{\rm P},\vartheta) \approx (0,0)$ (yellow). For each test particle, we accumulated $N_{\rm sec} = 900$ crossings. (b): Same Poincar\'{e} data as in (a) but plotted in the $(\hat{P}_\zeta,\vartheta)$-plane, where unperturbed orbit surfaces appear as straight horizontal lines, and an island with poloidicity $p=1$ appears at the seed resonance $\hat{P}_{\zeta,{\rm res}} \approx 1.325$ (dashed blue line) near the plasma center. The blue arrows indicate the sheared flow of off-resonant particles in the moving frame ($\zeta' = \zeta - \omega_0 t/n$). The resonant island width is determined by the fixed mode amplitude $A_0$, whose values (A) $10^{-3}$, (B) $2\times 10^{-3}$ and (C) $10^{-2}$ approximately represent the relative saturation levels in the respective simulation with active markers in Fig.~\protect\ref{fig:04_scanr} below.}%\vspace{-0.2cm}
	\label{fig:03_poink}%
\end{figure}

Fig.~\ref{fig:02_setup}(b) shows the radial dependence of the unperturbed guiding center resonant frequency $\Omega_{p,n}$ for fixed $K$ and low-field side pitch $u_{\rm lfs}/v$. Our guiding centers will interact with modes (A)--(C) only via the $p/n = 1/1$ resonance.\footnote{From the central core to the edge, the characteristic guiding center drift orbit frequencies lie in the range $\nu_{\rm tor} \approx (220...210)\,{\rm kHz}$, $\nu_{\rm pol} \approx (180...100)\,{\rm kHz}$, so the $p=0$ and $p=2$ resonances are too far away to play a role here.}
Although our test modes (A)--(C) consist only of a single poloidal Fourier harmonic $m=2$ in Boozer coordinates, one can anticipate from Fig.~\ref{fig:02_setup}(h)--(j) that the mode projected onto the orbit contour has an orbit-based poloidal Fourier spectrum that contains multiple components including $m_{\rm orb} = 1$ that facilitates a resonance with poloidicity $p=1$.\footnote{The poloidicity $p$ of the dominant resonance can be estimated by mapping the 2-D mode structure $\delta\tilde{\Phi}(R,z) = \delta\tilde{\Phi}(r) e^{im\vartheta}$ onto drift orbits. For long-wavelength (low-$n$) modes interacting with passing fast ions, it is often possible to identify a unique dominant resonance with effective orbit-based poloidal mode number $p = m_{\rm orb}$. Note that the $m$-spectrum depends on the chosen metric of $\vartheta$, whereas $m_{\rm orb}$ does not since it is measured in a drift orbit's time domain. See Appendix~\protect\ref{apdx:res} and Ref.~\protect\cite{Bierwage14} for details.}

As illustrated in Fig.~\ref{fig:02_setup}(b), the dynamically accessible range of frequencies in our simulation will be about $40...100\,{\rm kHz}$ for resonances ranging from the stagnation orbit near the axis to the plasma boundary. However, unlike in Fig.~\ref{fig:02_setup}(b), where we plotted the resonant frequency $\Omega_{p,n}$ for a fixed kinetic energy $K = 400\,{\rm keV}$, the simulation particles and the phase space dynamics they represent will approximately proceed along the line of constant total guiding center energy $\E'$ in the frame of reference rotating with phase velocity $\dot{\zeta} = \omega_0/n$,
\begin{equation}
	\E' = \E - \omega_0 P_\zeta/n \quad \text{(``rotating frame energy'')}.
	\label{eq:etot}
\end{equation}

\noindent (Systematic long-term deviations were measured in Fig.~29 of Ref.~\cite{Bierwage21}). Here, $P_\zeta$ is the canonical toroidal angular momentum and $\E$ is the total energy (Hamiltonian) of a guiding center with magnetic moment $\mu$ and parallel velocity $u$:
\begin{subequations}
	\begin{align}
		P_\zeta(t) =&\, -Ze\Psi_{\rm P,gc}(t) + M u(t) g(\Psi_{\rm P,gc}(t)),
		\label{eq:com_pzeta}
		\\
		\E(t) =&\, M u^2(t)/2 + \mu B_{\rm gc}(t) + Ze\delta\Phi({\bm X}_{\rm gc}(t),t),
		\label{eq:com_etot}
	\end{align}
	\label{eq:com}
\end{subequations}\vspace{-0.5cm}

\noindent with short-hand notation $B_{\rm gc}(t) \equiv B({\bm X}_{\rm gc}(t)$, etc.

The constraint in Eq.~(\ref{eq:etot}) is accounted for when we load simulation particles. Here, we use the same loading method (quiet start) as described in Section 3.3 of Ref.~\cite{Bierwage21}. At $t=0$, where $\delta\Phi = 0$, Eq.~(\ref{eq:etot}) with the constraint $\E' = {\rm const}$.\ reduces to
\begin{equation}
	\E'(0) = K(0) - \omega_0 P_\zeta(0)/n = {\rm const}.
	\label{eq:etot0}
\end{equation}

\noindent The magnetic moment is fixed at $\mu B_0 = 167.5\,{\rm keV}$ and we let $K_{\rm min} = 335\,{\rm keV}$ near the outer edge of the outer midplane, $(\psi_{\rm P}, \vartheta) \approx (1, 0)$. With the initial kinetic energies $K(0) = M u^2(0)/2 + \mu B(0)$ at each initial position $P_\zeta(0)$ obeying Eq.~(\ref{eq:etot0}), this yields $K_{\rm max} \approx 410\,{\rm keV}$ near the magnetic axis, as the color bar in Fig.~\ref{fig:03_poink} shows.

The triplet ${\bm C} = \{P_\zeta(0),\mu,\sigma(0) K(0)\}$ comprises one possible set of unperturbed constants of motion that uniquely define a toroidal drift orbit surface in the ambient field ${\bm B}$. In our visualizations of simulation results, we will use the normalization $\hat{P}_\zeta \equiv P_\zeta/(Ze\Psi_{\rm P,edge})$.

\begin{figure*}
	[tbp]\vspace{-0.45cm}
	\centering
	\includegraphics[width=0.96\textwidth]{\figures/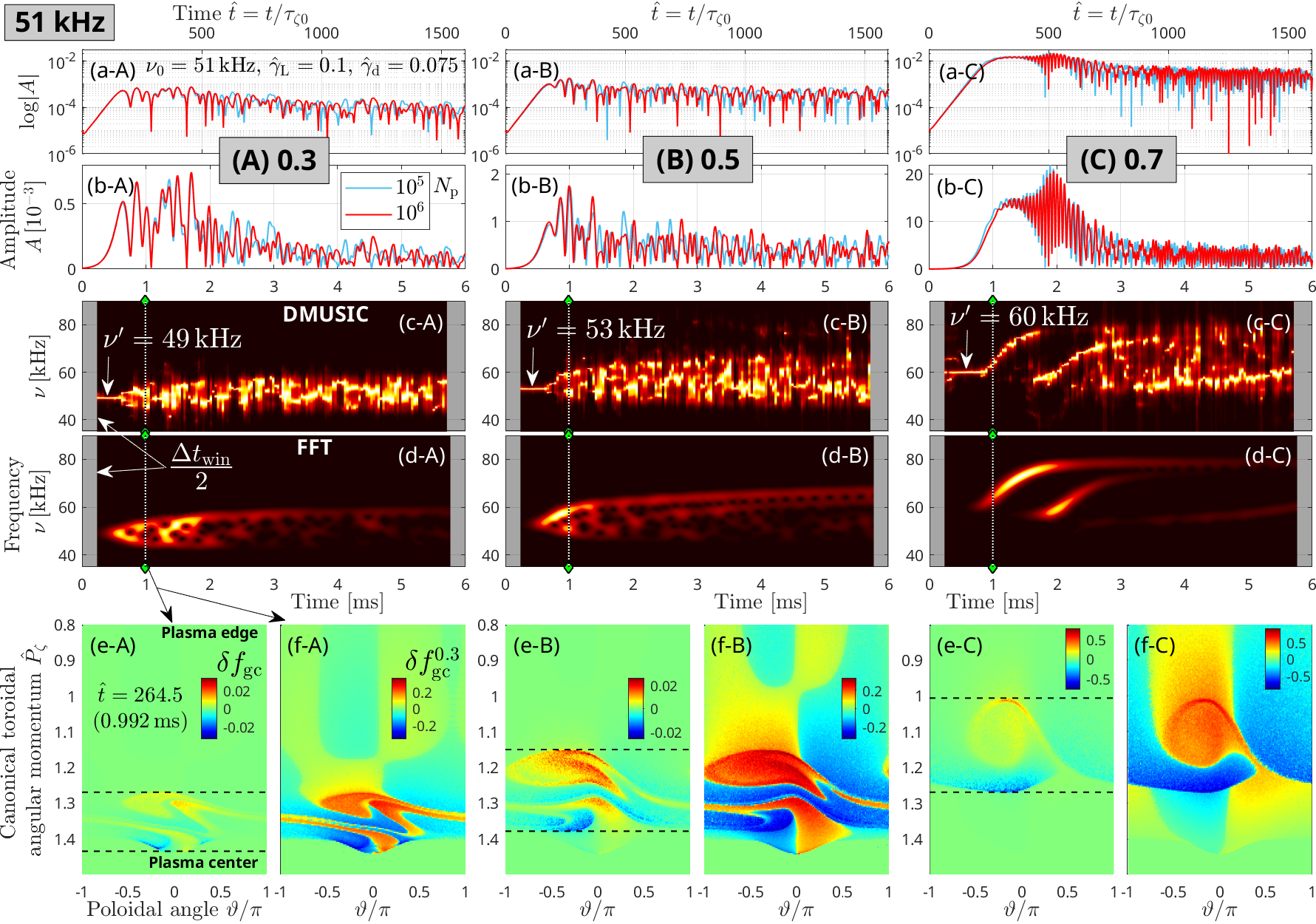}\vspace{-0.1cm}
	\caption{Overview of results obtained for the three cases of Fig.~\protect\ref{fig:02_setup}, where the mode's peak is located at (A) $\rpeak/a = 0.3$, (B) $0.5$, and (C) $0.7$. The seed frequency is $\omega_0 \approx 50.6\,{\rm kHz}$, the damping rate is $\hat{\gamma}_{\rm d} \equiv \gamma_{\rm d}/\omega_0 = 0.075$, and the gradients are chosen such that the undamped linear growth rate is $\hat{\gamma}_{\rm L} \approx 0.1$, so the net growth rate is $\hat{\gamma}_0 = \hat{\gamma}_{\rm L} - \hat{\gamma}_{\rm d} \approx 0.025$. (a,b): Evolution of the mode amplitude $A(t)$ on a logarithmic and linear scale. Times are given both in milliseconds (ms) and in units of the characteristic toroidal transit time $\tau_{\zeta 0} = 3.75\,\mu{\rm s}$ ($266.66\,{\rm kHz}$) defined in Eq.~(25) of Ref.~\protect\cite{Bierwage21}. The main simulations were performed using $N_{\rm p} = 10^6$ particles (red). Very similar results, especially during the first $2\,{\rm ms}$, were obtained with $10^5$ particles (light blue), and with a $4$ times smaller Runge-Kutta time step (not shown). The initial mode amplitude was chosen to be $A_0 \equiv A(0) = 10^{-5}$ and very similar results were obtained with $A_0 = 10^{-10}$ (not shown). (c,d): Spectrograms obtained using {\tt DMUSIC} with time window $\Delta t_{\rm win} \approx 0.375\,{\rm ms}$ (upper) and {\tt FFT} with zero-padded Hann-weighted time window $\Delta t_{\rm win} \approx 0.47\,{\rm ms}$ (lower). (e): Poincar\'{e} contour plots weighted by the phase space density perturbation $\delta f_{\rm gc}(P_\zeta,\vartheta)$ in the toroidally rotating plane $\zeta' = \zeta  - \omega_0 t/n$. The snapshot time is $t = 264.5\,\tau_{\zeta 0} \approx 1\,{\rm ms}$. Horizontal black dashed lines indicate the locations of the perturbation's lower (radially inward-propagating) and upper (radially outward-propagating) fronts. (f): Same data as in (e) but plotted as $|\delta f_{\rm gc}|^{0.3} \times \delta f_{\rm gc}/|\delta f_{\rm gc}|$ to enhance the contrast and highlight finer details of the phase space structures.}\vspace{-0.15cm}
	\label{fig:04_scanr}%
\end{figure*}

The Poincar\'{e} plots in Fig.~\ref{fig:03_poink} show the domain that has been loaded with simulation particles following the above recipe, the structure of the seed resonance for fixed mode amplitude and phase ($A,\phi = {\rm const}.$), and how the perturbed guiding center orbits of passive test particles overlap with the respective mode in cases (A)--(C).

The simulations discussed in the following Section~\ref{sec:results} were performed with $N_{\rm p} = 10^5$ or $10^6$ active marker particles, which are expected to stay near the initially loaded line defined by Eq.~(\ref{eq:etot}) for at least a few milliseconds (cf.~Figs.~C.1 and C.2 of Ref.~\cite{Bierwage21}). The characteristic toroidal transit period $\tau_{\zeta 0} \approx 2\pi R_0/\sqrt{2 K_0/M} \approx 3.75\,\mu{\rm s}$ ($1/\tau_{\zeta 0} \approx 266.66\,{\rm kHz}$) is used as an alternative unit of time (see Eq.~(25) of Ref.~\protect\cite{Bierwage21} for the precise definition).\vspace{-0.3cm}

%==============================================================================
\section{Simulation results}\vspace{-0.25cm}
\label{sec:results}

\begin{figure}
	[tbp]\vspace{-0.4cm}
	\centering
	\includegraphics[width=0.48\textwidth]{\figures/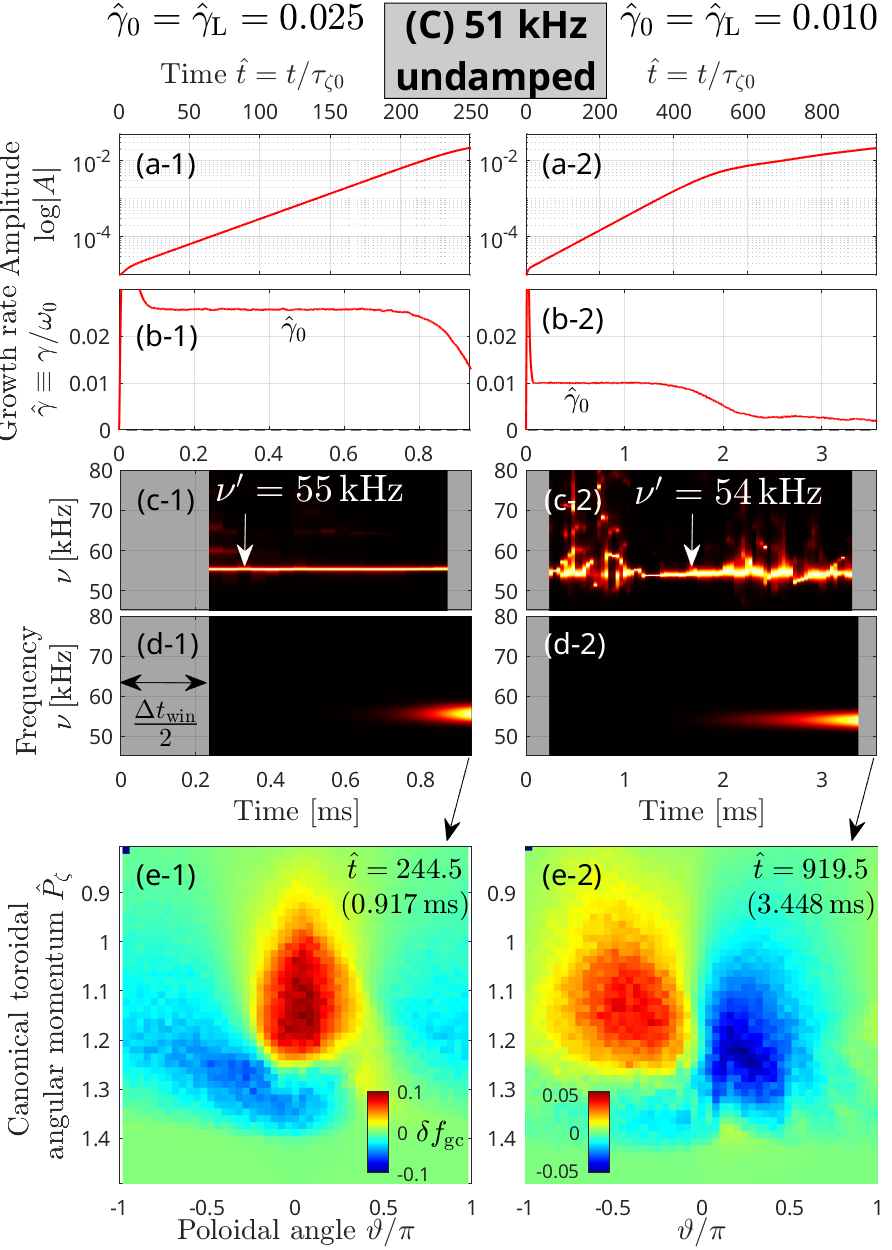}\vspace{-0.2cm}
	\caption{Results for the undamped case (C) with mode peak radius $\rpeak/a = 0.7$ for two different values of the gradient $\partial f_{\rm gc,0}/\partial{\bm C}$ giving the default net growth rate $\hat{\gamma}_0 = 0.025$ (left) and a reduced $\hat{\gamma}_0 = 0.010$ (right). Arranged as Fig.~\protect\ref{fig:04_scanr}, except that panel group (b) shows the instantaneous growth rate $\hat{\gamma}(t)$ panel group (f) is omitted, and the histograms of Poincar\'{e} crossings in row (e) have only $50$ (instead of $200$) diagnostic cells in both dimensions. These simulations were not suitably prepared for long run times with high amplitudes, so we terminated them when $A(t)$ reached $2\times 10^{-2}$.}\vspace{-0.4cm}
	\label{fig:05_damp0}%
\end{figure}

When the seed resonance's drift orbits lie mostly on one side of the mode's peak, we found previously that the wave raises or drops its angular phase velocity by $\delta\omega_0 = \dot{\phi}$ in such a way that it resonates with drift orbits whose mean radius is closer to the mode's peak radius $\rpeak$. That is, the resonance moves to (or reforms at) a different location ${\bm C}' = {\bm C}+\delta{\bm C}$ in the space of unperturbed constants of motion ${\bm C} = \{P_\zeta(0),\mu,\sigma(0)K(0)\}$, satisfying\vspace{-0.15cm}
\begin{equation}
	\omega' = \Omega_{p,n}({\bm C}') \quad \text{with} \quad \omega' \equiv \omega_0 + \delta\omega_0,
	\label{eq:res3}
\end{equation}

\noindent where $\Omega_{p,n}({\bm C}_0)$ is the local resonant orbit frequency around the radius $\rpeak$ of the mode's peak. As its name implies, the prompt shift $\delta\omega_0 = \dot{\phi} = {\rm const}$.\ is established almost instantaneously, here on the time scale of half an oscillation period $\pi/\omega_0$, which corresponds to a few toroidal transit periods $\tau_\zeta$ of resonant particles (see Fig.~\ref{fig:05_damp0}(b) below or, for more detail, Fig.~D6(b) of Ref.~\cite{Bierwage21}). While $\delta\omega_0$ is subsequently independent of the mode amplitude $A(t)$, its sign $\delta\omega_0/|\delta\omega_0|$ is such that the resonance is shifted towards the mode's peak, and its magnitude $|\delta\omega_0|$ increases with increasing mismatch $|\omega_{\rm opt} - \omega_0| \approx |\Omega_{p,n}({\bm C}_0) - \omega_0|$, where $\omega_{\rm opt}$ is a presumed but unknown ``optimized'' frequency that in the present setup seems to be close to $|\Omega_{p,n}({\bm C}_0)$ at the mode's peak.

In Fig.~D6 of Ref.~\cite{Bierwage21}, it was shown that the sign and magnitude of the prompt frequency shift $\delta\omega_0 = 2\pi\delta\nu_0$ depend not only on the relative location of the mode's peak and the seed resonance, but also has a weak dependence on the rate of dissipative damping $\gamma_{\rm d}$ that, together with the linear drive $\gamma_{\rm L}$, determines the net linear growth rate $\gamma_0 = \gamma_{\rm L} - \gamma_{\rm d}$. In the following Sections~\ref{sec:results_scanr} and \ref{sec:results_damp0}, we revisit these dependencies in the present JT-60U-based scenario. We will also examine the ensuing nonlinear dynamics. In the final Section~\ref{sec:results_scanf}, we present key results of scanning the seed frequency $\omega_0 = 2\pi\nu_0$ in case (C) with $\rpeak/a = 0.7$.

%------------------------------------------------------------------------------
\subsection{Scan of mode peak location $\rpeak$}
\label{sec:results_scanr}

Fig.~\ref{fig:04_scanr} shows an overview of the main results obtained for the three cases of Fig.~\protect\ref{fig:02_setup}, where we varied the location $\rpeak$ of the mode's peak. Within the first $\hat{t} \lesssim 3$ transit times ($\hat{t} \equiv t/\tau_{\zeta 0}$), we observe a prompt frequency shift by $\delta\nu_0 \equiv \delta\omega_0/(2\pi)$:
\begin{equation}
	\nu_0 = 50.6,{\rm kHz} \rightarrow \nu' = \nu_0 + \delta\nu_0 \approx \{49,53,60\}\,{\rm kHz}.
\end{equation}

\noindent The values of $\nu'$, listed in $\{...\}$ brackets for cases \{(A), (B), (C)\}, were inferred from DMUSIC\footnote{See Appendix E.2 of Ref.~\protect\cite{Bierwage21} and the original paper \protect\cite{LiY98}.}
spectrograms in row (c) of Fig.~\ref{fig:04_scanr}. The main features of the subsequent nonlinear chirping dynamics are perhaps best seen in the FFT spectrograms that are plotted in row (d).

For mode (A) with $\rpeak/a = 0.3$, we observe in Fig.~\ref{fig:04_scanr}(c-A) a slightly down-shifted $\nu' \approx 49\,{\rm kHz}$ ($\delta\nu_0 \approx -1.5\,{\rm kHz}$). The subsequent chirp during $0.5\,{\rm ms} \lesssim t \lesssim 1\,{\rm ms}$ in Fig.~\ref{fig:04_scanr}(d-A) is predominantly downward (radially inward), suggesting that mode (A) is affected most strongly by orbits closer to the stagnation point near the plasma center. However, being already close to the center, the downward chirp in Fig.~\ref{fig:04_scanr}(d-A) soon terminates, so that upward chirping (radially outward) dominates in the long run. The mode amplitude $A(t)$ in panels (a-A) and (b-A) exhibits the usual beating dynamics \cite{Bierwage21}, reaching peak values around $7\times 10^{-4}$ during the first $2\,{\rm ms}$, and subsequently decays gradually.

For mode (B) with $\rpeak/a = 0.5$, we observe in Fig.~\ref{fig:04_scanr}(c-B) a slightly up-shifted $\nu' \approx 53\,{\rm kHz}$ ($\delta\nu_0 \approx +2.5\,{\rm kHz}$). The subsequent chirp in Fig.~\ref{fig:04_scanr}(d-B) is predominantly upward, consistently with the fact that the mode's peak in Fig.~\ref{fig:03_poink}(a-B) is effectively located outside the seed resonance radius. The mode amplitude $A(t)$ in Fig.~\ref{fig:04_scanr}(a-B) and (b-B) seems to initially oscillate more or less harmonically before starting to beat in an increasingly chaotic manner. It reaches peak values around $1.6\times 10^{-3}$ during the first $2\,{\rm ms}$ and subsequently decays gradually.

For mode (C) with $\rpeak/a = 0.7$, we observe in Fig.~\ref{fig:04_scanr}(c-C) a significantly up-shifted $\nu' \approx 60\,{\rm kHz}$ ($\delta\nu_0 \approx +10\,{\rm kHz}$). The subsequent chirp in Fig.~\ref{fig:04_scanr}(d-C) is upward, consistently with the fact that the mode's peak in Fig.~\ref{fig:03_poink}(a-C) is located far outside the resonant radius. The mode amplitude $A(t)$ in Fig.~\ref{fig:04_scanr}(a-C) and (b-C) first saturates around $\hat{t} \approx 1\,{\rm ms}$ at a level of $15\times 10^{-3}$. This is followed by a burst of rapid pulsations, during which $A(t)$ reaches about $20\times 10^{-3}$. The burst lasts a few $100\,\mu{\rm s}$, peaks around $t \approx 1.9\,{\rm ms}$, and coincides with the appearance of a second spectral peak near $60\,{\rm kHz}$ in Fig.~\ref{fig:04_scanr}(d-C). The pulsations begin with a frequency of about $1/(0.05\,{\rm ms}) = 20\,{\rm kHz}$ and slow down afterwards, which is consistent with the beating between the primary spectral peak, now staying near $80\,{\rm kHz}$, and the secondary peak chirping upward from $60\,{\rm kHz}$.

Overall, we observe that the effect of the prompt frequency shift persists in the nonlinear regime ($t > 0.5\,{\rm ms}$) in all cases in Fig.~\ref{fig:04_scanr}(c,d) as the fluctuating spectral bands are progressively shifted towards higher frequencies from (A) to (C). The phase space structures in Fig.~\ref{fig:04_scanr}(e,f) become increasingly plume-like, with radial propagation ($\dot{P}_\zeta$) becoming increasingly rapid compared to poloidal propagation ($\dot{\vartheta}$) and shearing ($\partial\dot{\vartheta}/\partial{P_\zeta}$).

In the extreme case (C), a giant vortical clump ($\delta f_{\rm gc} > 0$) balloons far up along the $\hat{P}_\zeta$ axis in Fig.~\ref{fig:04_scanr}(e,f-C), without even fully detaching from its point of origin on the other side of the seed resonance. Moreover, this plume-like clump vortex exhibits a large internal nonuniformity, with a $\delta f_{\rm gc}$ spike forming near its upper front ($\hat{P}_\zeta \approx 1.0$). Comparing the radial extent $1.0 \lesssim \hat{P}_\zeta \lesssim 1.2$ of the giant clump in Fig.~\ref{fig:04_scanr}(f-C) with the Poincar\'{e} contours in Fig.~\ref{fig:03_poink}(a,b-C) tells us that it overlaps well with the mode's peak whenever it passes the outer midplane ($\vartheta \approx 0$).

\begin{figure*}
	[tbp]\vspace{-0.7cm}
	\centering%\vspace{-0.05cm}
	\includegraphics[width=0.96\textwidth]{\figures/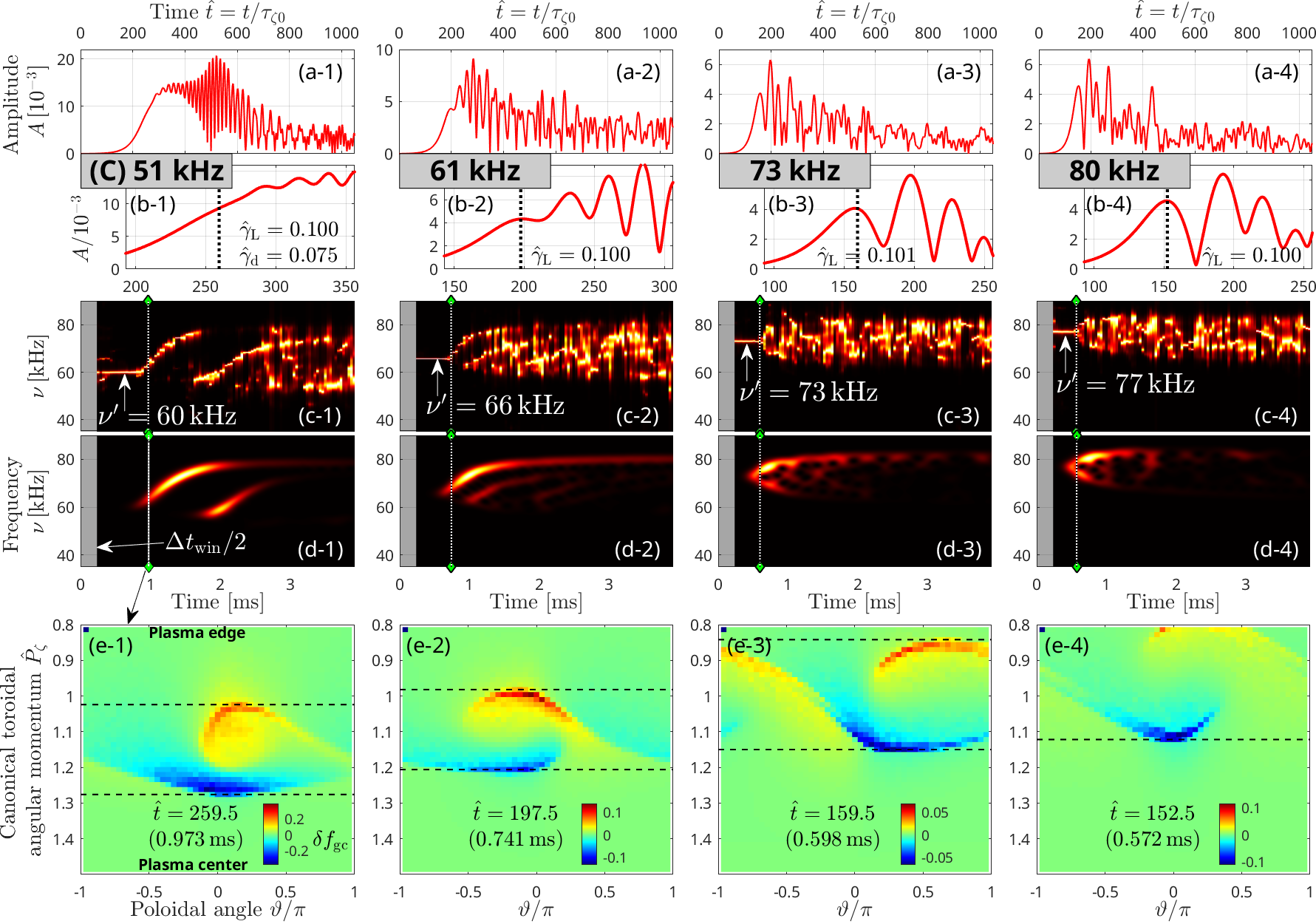}\vspace{-0.15cm}
	\caption{Overview of results of a seed frequency scan in case (C) with mode peak radius $\rpeak/a = 0.7$. Arranged as Fig.~\protect\ref{fig:04_scanr}, except that panel group (a) has a linear scale and (b) shows an enlargement thereof. Moreover, as in Fig.~\protect\ref{fig:05_damp0}, (e) uses a coarser diagnostic grid, and (f) is omitted. Each column summarizes data for a different seed frequency $\nu_0 \approx (51,\, 61,\, 73,\,80)\,{\rm kHz}$.}\vspace{-0.25cm}
	\label{fig:06_scanf}%
\end{figure*}

In part, this plume-like behavior can be attributed to the fact that phase space structure propagating radially outward (upward in our Poincar\'{e} plots, towards smaller values of $\hat{P}_\zeta$) experience (pre)amplified fluctuation amplitudes, which increases also the effective radial (vertical) width $\Delta\hat{P}_{\zeta,{\rm res}}$ of the red-shaded clump structure. The fact that the seed resonance in case (C) lies beyond the inflection point of the Gaussian mode profile\footnote{To see this, compare Fig.~\ref{fig:04_scanr}(e,f-C) with Fig.~\ref{fig:03_poink}(a,b-C).}
likely contributes since the fluctuating field intensity increases at an increasing rate until the inflection point is reached.\footnote{In Ref.~\cite{Bierwage21}, we dubbed this effect ``pre-convective amplification''. It may occur when a mode's radial profile is relatively robust.}

It must be noted that we have prepared cases (A)--(C) in Fig.~\ref{fig:04_scanr} such that they all have the same linear growth rate $\hat{\gamma}_0 \equiv \gamma_0/\omega_0 \approx 0.025$ by fixing the rates of damping $\hat{\gamma}_{\rm d} = 0.075$ and linear drive $\hat{\gamma}_{\rm L} \approx 0.1$. In order to achieve this, the gradients $\partial \ln f_{\rm gc,0}/\partial{\bm C}$ had to be increased from one case to the next. The gradient ratios for (A):(B):(C) are $0.027:0.06:1$ or, equivalently, $1:2.22:37.0$. Since the gradients are initially taken to be uniform across the entire simulation domain, the mode in case (C) can feed on a much larger reservoir of free energy than (A) and (B). Of course, mode (C) also occupies a larger volume, but scaling the above ratios by the respective values of $a/\rpeak$, still yields appreciable ratios $0.063:0.084:1$ or, equivalently, $1:1.33:15.9$. The latter are consistent with the relative ratios $0.7:1.6:15$ of the peak amplitudes seen in Fig.~\ref{fig:04_scanr}(a,b).

%------------------------------------------------------------------------------
\subsection{Role of gradients and damping for $\rpeak/a = 0.7$}\vspace{-0.15cm}
\label{sec:results_damp0}

Most of the above gradient scaling factor for case (C) was needed to raise $\hat{\gamma}_{\rm L} \equiv \gamma_{\rm L}/\omega_0$ above the damping rate $\hat{\gamma}_{\rm d} = 0.075$, such that the mode could grow at the net rate of $\hat{\gamma}_0 = \hat{\gamma}_{\rm L} - \hat{\gamma}_{\rm d} = 0.025$ (our default choice).

Reducing the gradient in case (C) by $13\%$ results in a marginally stable configuration with $\hat{\gamma}_0 = \hat{\gamma}_{\rm L} - \hat{\gamma}_{\rm d} \approx 0$. Meanwhile, we find that the promptly shifted frequency remains unchanged at $\nu' = \nu_0 +\delta\nu_0 \approx 60\,{\rm kHz}$. Since the mode does not grow, the resonance in this marginally stable case (not shown here) becomes very sharp after about $200\,\tau_{\zeta 0}$ toroidal transits ($0.75\,{\rm ms}$) via phase mixing, as was shown in Fig.~11 of Ref.~\cite{Bierwage21}. This allowed us to determine with high precision the location of the shifted resonance: $\hat{P}_{\zeta,{\rm res}}(60\,{\rm kHz}) \approx 1.205$. This is near the bottom rim of the giant clump in Fig.~\ref{fig:04_scanr}(f-C). For comparison, the seed resonance is located at $\hat{P}_{\zeta,{\rm res}}(51\,{\rm kHz}) \approx 1.325$ as we saw in Fig.~\ref{fig:03_poink}(b).

In the absence of damping ($\hat{\gamma}_{\rm d} = 0$), the default net growth rate $\hat{\gamma}_0 = \hat{\gamma}_{\rm L} = 0.025$ is obtained when the gradient is reduced by $33\%$ below its default value. $50\%$ of the default gradient gives $\hat{\gamma}_0 = \hat{\gamma}_{\rm L} = 0.010$. The results for these two undamped variants of case (C) are summarized in Fig.~\ref{fig:05_damp0}, where one can see that reducing $\hat{\gamma}_{\rm d}$ from $0.075$ to $0$ reduces the promptly shifted frequency $\nu' = \nu_0 +\delta\nu_0$ by approximately $10\%$, from $60\,{\rm kHz}$ to $54...55\,{\rm kHz}$.

%------------------------------------------------------------------------------
\subsection{Scan of the seed frequency $\nu_0$ for $\rpeak/a = 0.7$}
\label{sec:results_scanf}

As the title ``auto-optimization of resonant drive'' of Appendix~D.4 in Ref.~\cite{Bierwage21} implies, we hypothesized previously that the increase of the prompt frequency shift $\delta\omega_0 = \dot{\phi}$ with increasing distance of the peak radius $\rpeak$ from the seed resonance as seen in Fig.~\ref{fig:04_scanr} would be such that the shifted frequency $\omega' = \omega_0 + \delta\omega_0$ allows the mode to take maximal advantage of the resonant drive that is available in its domain of influence.

To test this hypothesis, we repeat the simulation with a new seed frequency near the promptly shifted frequency of the previous run: $\nu_0 \rightarrow \nu_0^{(1)} = \nu_0 + \delta\nu_0 \approx 60\,{\rm kHz}$. If this was the optimal configuration as hypothesized, then no further prompt frequency shift should occur ($\dot{\phi} = 0$). However, this is not what we find. Instead, the second column of Fig.~\ref{fig:06_scanf} shows that for $\nu_0^{(1)} = 61\,{\rm kHz}$ the frequency promptly shifts to a yet higher value of about $\nu' = \nu_0^{(1)} + \delta\nu_0^{(1)} \approx 67\,{\rm kHz}$. Iterating this procedure as
\begin{equation}
	\nu_0^{(k)} = \nu_0^{(k-1)} + \delta\nu_0^{(k-1)},
	\label{eq:nu0_iterate}
\end{equation}

\noindent we find that the prompt frequency shift vanishes for $\nu_0 \approx 73\,{\rm kHz}$, as can be seen in the third column of Fig.~\ref{fig:06_scanf}. Going beyond this value, yields prompt frequency shifts in the opposite direction, as shown in the last column of Fig.~\ref{fig:06_scanf}, where $\nu_0 = 80\,{\rm kHz}$ gives $\nu' = \nu_0 + \delta\nu_0 = 77\,{\rm kHz}$.

Comparability of the results of the seed frequency scan in Fig.~\ref{fig:06_scanf} was ensured by performing all simulations with the same set of simulation particles, which were loaded using Eq.~(\ref{eq:etot}) with $\omega_0 = 2\pi\times 50.6\,{\rm kHz}$. Nevertheless, we are aware of two minor systematic caveats in our setup.

First, the ideal MHD constraint that was mentioned in the first paragraph of Section~\ref{sec:model} is satisfied only by the seed wave component in Eq.~(\ref{eq:mode}), because our model assumes that $\nabla_\parallel\delta\Phi = i\omega_0\alpha B$. Hence, in accordance with Faraday's law, which is built into the guiding center Lagrangian underlying {\tt ORBIT}'s equation of motion, there is a parallel electric field $\delta E_\parallel$ proportional to the instantaneous phase drift $\dot{\phi}(t) = i\partial_t - \omega_0$. Test runs with a modified version of {\tt ORBIT}, where instances of $i\omega_0\alpha$ were replaced by $-\partial_t\alpha = (i\omega_0 - \dot{\phi})\alpha$, showed only minor quantitative differences that we considered to be ignorable for the purposes of the present paper.

Second, since we fixed the normalized values $\hat{\gamma}_{\rm d} \equiv \gamma_{\rm d}/\nu_0 = 0.075$ and $\hat{\gamma}_{\rm L} \equiv \gamma_{\rm L}/\nu_0 \approx 0.1$, the actual damping and growth rates $\gamma_{\rm d}$ and $\gamma_{\rm L}$ in the numerator increased together with $\nu_0$ in the denominator. In order to assess the impact of this caveat, we performed another simulation with $\nu_0^{(1)} = 61\,{\rm kHz}$ using rescaled $\hat{\gamma}_{\rm d}^{(1)} \equiv \hat{\gamma}_{\rm d} \nu_0/\nu_0^{(1)} \approx 0.0622$ and $\hat{\gamma}_{\rm L}^{(1)} \equiv \hat{\gamma}_{\rm L} \hat{\gamma}_{\rm d}^{(1)}/\hat{\gamma}_{\rm d} \approx 0.0830$. (The gradients $\partial\ln f_{\rm gc,0}/\partial{\bm C}$ were slightly increased by a factor $1.05$ to obtain the desired value of $\gamma_{\rm L}^{(1)}$.)
The resulting promptly shifted frequency $\nu_0^{(1)} + \delta\nu_0^{(1)} \approx 66\,{\rm kHz}$ was still very close to the $67\,{\rm kHz}$ measured in Fig.~\ref{fig:06_scanf}(c-2). We thus consider the frequency scan in Fig.~\ref{fig:06_scanf} to be meaningful.\vspace{-0.2cm}

%------------------------------------------------------------------------------
\subsection{Discussion of incomplete auto-optimization}
\label{sec:results_disc}\vspace{-0.1cm}

In Section~\protect\ref{sec:results_damp0}, we verified that changing the drive while keeping all other parameters constant did not affect the prompt frequency shift significantly. We took this as a justification for preparing all cases in Fig.~\ref{fig:06_scanf} to have the same linear driving rate $\hat\gamma_{\rm L} \approx 0.1$. This was realized by adjusting the gradients in the second, third and fourth column of Fig.~\ref{fig:06_scanf} by scaling factors $0.44$, $0.30$ and $0.32$, respectively. These values tell us that the $73\,{\rm kHz}$ case in the third column of Fig.~\ref{fig:06_scanf} (without prompt frequency shift) required the least drive to obtain the desired growth rate. In other words,
\begin{itemize}
	\item  the seed frequency that minimizes the prompt frequency shift also maximizes the resonant drive efficiency (or energy conversion rate).
\end{itemize}

\noindent Since the prompt frequency shift $\delta\omega_0$ has lead us towards the point where the energy conversion rate is maximized, $\delta\omega_0$ can be regarded as being the manifestation of an intrinsic prompt auto-optimization process, which seems to have partly survived the time-scale-separation-based truncation of rapidly varying terms in the unfiltered Eq.~(\ref{eq:pert_A_phi_unfilt}), as we discussed in Section~\ref{sec:model}.

While that imperfection of the time-scale-separation procedure is the most likely explanation, it is worth considering and ruling out other factors:
\begin{itemize}
	\item {\it Boundary conditions:} Particles have been loaded in a sufficiently wide range of energies and canonical momenta (obeying $\E' = {\rm const}$.) to make the optimal resonance promptly accessible in principle.
	
	\item {\it Limited resonance width:} One may wonder whether the prompt frequency shift is incomplete because the distance between the seed and the target resonance exceeds the resonance width measured, for instance, by $\gamma = 0.025\omega_0 \approx 10^4\,{\rm s}^{-1}$. Several arguments speak against this idea. First, we do not observe any threshold: the prompt frequency shift always falls short of complete auto-optimization even when we approach the target asymptotically. Second, Fig.~\ref{fig:05_damp0}(b) shows that the effective growth rate is much larger during the short initial interval $0 \leq t \lesssim \pi/\omega_0 \approx 10\,\mu{\rm s}$ during which the prompt frequency shift takes place. Indeed, the frequency uncertainty during this short interval is at least $\nu_0 \pm 2/\tau_0 \sim (0...100)\,{\rm kHz}$, so the optimal resonant frequency $73\,{\rm kHz}$ found in Fig.~\ref{fig:06_scanf} lies well within the expected accessible range.
	
	\item {\it Limited orbit width:} In Fig.~\ref{fig:03_poink}(a-C), the radial excursion of the seed resonance due to magnetic drifts does not reach all the way to the mode's peak. While this may contribute to the incompleteness of the prompt frequency shift, it is not a necessary factor because iterations (\ref{eq:nu0_iterate}) are still necessary after $\rpeak$ has come within reach of the orbit width.
\end{itemize}

\noindent The observed frequency shifts thus fall far short of the expected physical possibilities. Therefore, we suspect that the auto-optimization in our perturbative model is incomplete because terms needed to fully describe short-time-scale  dynamics ($\ddot{A}$, $\ddot{\phi}$, $\dot{\phi}\dot{A}$, etc.) were truncated in the derivation of Eq.~(\ref{eq:pert_A_phi_unfilt}). We suspect that the prompt frequency shift is reduced further if one enforces the slowness condition (IV) in Section~\ref{sec:model} via a low-pass filter, similar to the $\tau_0$-average in Eq.~(\ref{eq:pert_A_phi}) that appears in Refs.~\cite{ChenY99,WhiteTokBook3} but was not actually implemented in {\tt ORBIT}.\vspace{-0.2cm}

%==============================================================================
\section{Summary and conclusion}\vspace{-0.2cm}
\label{sec:summary}

In this paper, we revisited the prompt shift $\delta\omega_0$ of the resonant frequency \cite{Bierwage21} that was observed at the beginning of simulations using the code {\tt ORBIT}. The version of the code used here employs a reduced perturbative model (Section~\ref{sec:model}) of fast-ion-driven Alfv\'{e}n modes to capture certain aspects of nonlinear chirping and associated fast ion transport in a tokamak plasma. Our interpretation in Appendix D.4 of Ref.~\cite{Bierwage21} had been that the prompt frequency shift occurs when the user-prescribed seed frequency $\omega_0$ of the symmetry-breaking initial perturbation does not coincide with the frequency of the resonance that maximizes the energy conversion rate between fast ion gradients (free energy source) and electromagnetic fluctuations (mode growth). Let us call this the ``auto-optimization hypothesis''.

In Section~\ref{sec:results}, we presented results of numerical experiments whose purpose was to test the said auto-optimization hypothesis by throwing more light on various parameter dependencies. The main findings are as follows:
\begin{itemize}
	\item  The prompt frequency shift $\delta\omega_0$ is indeed directed towards the resonance at which the drive efficiency (or energy conversion rate is) maximized.
	\item  However, $\delta\omega_0$ is always too small to reach the optimal target frequency. Iteration as in Eq.~(\ref{eq:nu0_iterate}) is required to asymptotically converge to the target.
	\item  In short, the auto-optimization in the present perturbative model is incomplete but can be completed iteratively with very small computational cost.
\end{itemize}

Indeed, as noted in Section~\ref{sec:model}, the present perturbative model was derived under the assumption of slowly varying amplitude and phase, ${\rm d}\phi/{\rm d}t \ll \omega_0$, ${\rm d}\ln A/{\rm d}t \ll \omega_0$ \cite{ChenY99, Pinches98}, so the time scales of mode structure formation and auto-optimization has formally been separated out. According to the universal uncertainty principle, the characteristic time scale for these processes in a comprehensive multi-time-scale model should be about half an oscillation period, $\tau_0/2 = \pi/\omega_0$. Incidentally, this coincides rather accurately with the time scale of the observed prompt frequency shift, so we may indeed interpret it as a remnant of the physical auto-optimization process.

Our finding is both interesting and concerning:
\begin{itemize}
	\item  On the one hand, it is remarkable that the prompt frequency shift occurs at all, even if only in rudimental form. We had not anticipated this. As noted in Section~\ref{sec:model}, this demonstrates that time-scale-separation-based orderings and truncations do {\it not} work like a perfect low-pass filter. No matter how rigorous the mathematical derivation had been, one should be cautious of remnants of formally ordered-out short-time-scale dynamics in numerical simulations based on such models.\footnote{When the fast dynamics depend on a certain degree of freedom, such remnants may be avoided by entirely eliminating that degree of freedom. For instance, the existence of fast magnetoacoustic waves relies on the magnetic fluctuations having a parallel component, $\delta B_\parallel \equiv \hat{\bm b}\cdot\delta{\bm B}$, which can be systematically eliminated (as in reduced MHD). Meanwhile, gyrokinetic simulations using models that retain portions of $\delta B_\parallel$ should be treated with care.}
		
	\item  On the other hand, the fact that transients like our the prompt frequency shift lie outside the model's domain of validity means that we cannot explain the origin of $\delta\omega_0$ with absolute certainty.
\end{itemize}

\noindent Nevertheless, the empirical evidence gathered from numerical experiments seems to support our claim that $\delta\omega_0$ is {\it an incomplete yet meaningful manifestation (or remnant) of auto-optimization of resonant drive}. The observed parameter-dependencies of $\delta\omega_0$ exhibit clear trends. We have also confirmed that $\delta\omega_0$ is independent of the initial perturbation amplitude and insensitive to numerical parameters (time step, particle number and loading method). It should also be noted that, once established, $\delta\omega_0$ is very robust and exerts a lasting influence on the subsequent nonlinear dynamics in the form of chirping and transport.

On this basis, we think that it is justified to consider potential applications. With the auto-optimization hypothesis confirmed --- albeit in incomplete form, requiring iteration (\ref{eq:nu0_iterate}) --- the prompt frequency shift could help us to infer the initial frequency of an energetic particle mode (EPM \cite{Chen94}) that is consistent with the fast ion drift orbits and their distribution in the space of unperturbed constants of motion ${\bm C}$. As we noted in the introductory Section \ref{sec:intro}, this would be an important step forward in efforts to capture EPM effects in computationally efficient reduced models and the integrated workflows they are embedded in. The fact that long-wavelength ($n=1$) EPMs were seen in beam driven JT-60U plasmas, motivated our choice to use the JT-60U scenario previously studied in Refs.~\cite{Bierwage13, Bierwage14, Bierwage16a, Bierwage16b, Bierwage17a} also in the present paper.

\begin{figure}
	[tbp!]\vspace{-0.5cm}
	\centering
	\includegraphics[width=0.48\textwidth]{\figures/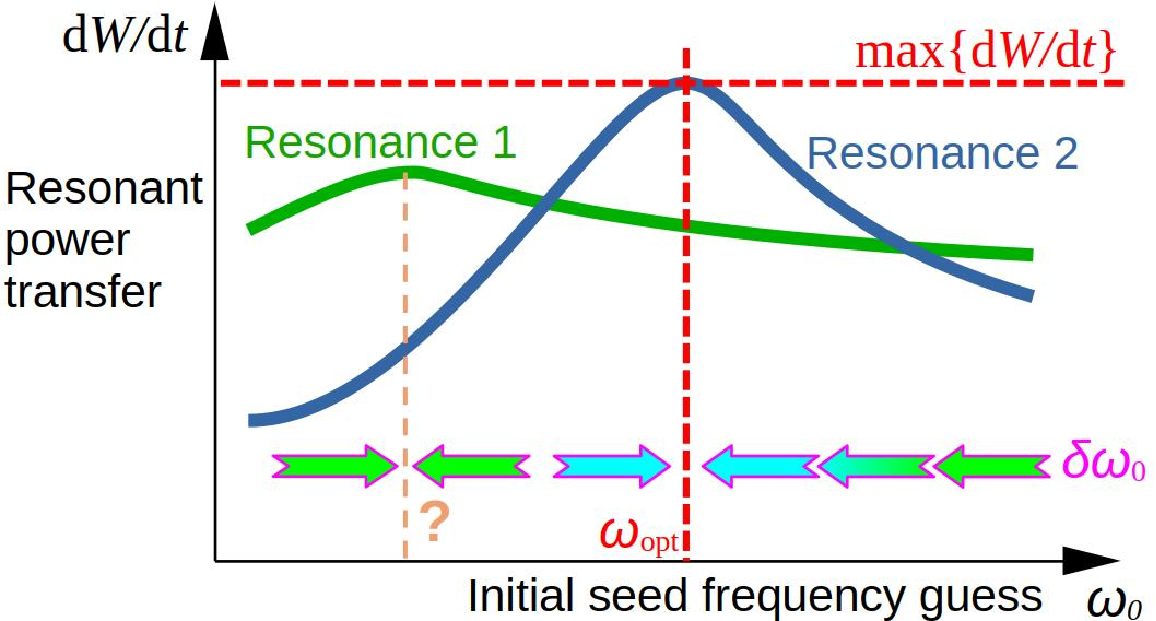}\vspace{-0.2cm}
	\caption{Schematic illustration of the possibility (to be confirmed) that the prompt frequency shift $\omega_0 \rightarrow \omega' = \omega_0 + \delta\omega_0$ associated with (incomplete) auto-optimization studied in the present paper could lead only to the nearest local maximum, like the one on the left-hand side. Therefore, we recommend to scan the initial guess of the seed frequency $\omega_0$ in order to find the optimum $\omega_{\rm opt}$, where the resonant power transfer ${\rm d}W/{\rm d}t$ from fast ion gradients to a mode is maximized absolutely. After that, $\delta\omega_0$ may be used for fine-tuning.}\vspace{-0.35cm}
	\label{fig:07_schematic-scan-w0}%
\end{figure}

The significance of the prompt frequency shift depends on the particular case at hand. Our empirical understanding of the shift allowed us to deliberately enhance it for emphasis and easier study in suitably designed numerical experiments, where $\delta\omega_0/\omega_0 \lesssim 20\%$. Without such deliberate enhancement, one may expect smaller shifts of a few \%. For some applications, this may be tolerable, so there may be no need to iteratively reinitialize the simulation with updated guesses of $\omega_0$ using Eq.~(\ref{eq:nu0_iterate}). Nevertheless, when using complete fast ion distributions that may drive multiple resonances simultaneously, we do recommend to begin with a scan of $\omega_0$ over the expected range of possible EPM frequencies (for instance, based on the structure of the continuous spectrum), because the optimal frequency $\omega_{\rm opt}$ may be obscured by other resonances in cases like that illustrated in Fig.~\ref{fig:07_schematic-scan-w0}.

Concerning further steps, we would like to make the following remarks:
\begin{itemize}
	\item  While the exaggerated Case (C) was useful for our study of the prompt frequency shift as such, Case (A) in Fig.~\ref{fig:04_scanr} is closest to the situation in comprehensive non-perturbative simulations that where validated against experiments \cite{Bierwage21}. The initial trend of downward chirping in Fig.~\ref{fig:04_scanr}(d-A) is correct. However, at later times, upward chirping dominates in the reduced perturbative model, while this is not seen in the non-perturbative simulation described in the above-mentioned references \cite{Bierwage13, Bierwage14, Bierwage16a, Bierwage16b, Bierwage17a}. Indeed, various tests have indicated that dominant downward chirping is difficult to obtain because, in the perturbative model, downward chirping is directed radially inward. Since our EPM of interest is already located in the central region of the plasma and since magnetic drifts are large for the relevant energies and pitch angles $(K,\mu)$, there is not much room for further inward propagation.
	
	\item  We attribute these qualitative discrepancies to the fact that the dominant chirps in non-perturbative simulations are guided by the continuous spectrum of shear Alfv\'{e}n waves. Thus, we believe that it is necessary to look for ways to include that information about the preferred plasma response in the reduced model. As was briefly mentioned in Section~\ref{sec:intro}, one idea is to incorporate information about continuous spectra in a radially nonuniform and frequency-dependent damping rate $\gamma_{\rm d}(\omega,r)$. Similarly, changes in the mode's spatial structure $\delta\Phi({\bm x})$ could also be modeled based on information about the continuous spectra (peak location) and magnetic drifts (radial width). The reduced simulations could be initialized with a set of test wavefunctions distributed along the continua.
		
	\item  The model used in this study assumed a uniform fast ion gradient ($\partial f_{\rm gc,0}/\partial P_\zeta = {\rm const}$.) and a single value of the magnetic moment. Simulation particles were loaded only along a thin sheet of phase space defined by Eq.~(\ref{eq:etot}) with $\E' = {\rm const}$.\ and $\Delta\E' \approx 0$. In the future, the proposed frequency-inference method should be tested in the presence of a more complete and more realistic fast ion distribution; in our case, at least in the energy range $300\,{\rm keV} \lesssim K \lesssim 400\,{\rm keV}$, where the dominant $n=1$ EPM resonance is located (see Fig.~8(h) of Ref.~\cite{Bierwage14}).
	
	\item  Since the prompt frequency shift can be measured after a few toroidal transit times in {\tt ORBIT}, the iteration (\ref{eq:nu0_iterate}) is expected to be computationally inexpensive. Nevertheless, it is worth investigating whether further speed-up is possible using approximate formulas like Eq.~(21) in Ref.~\cite{WeiG25}.
	
	\item  Fishbone-like EPMs and other low-frequency modes were found to have mixed polarization with a significant compressible component \cite{Du24}. Thus, before applying our model to such modes, the approximation $\delta{\bm B} \approx \nablab\times\alpha{\bm B}$ needs to be generalized.
\end{itemize}

%==============================================================================
\begin{acknowledgments}
A.B.\ thanks Kawamura Gakushi, Shimpei Futatani (QST), Philipp Lauber (MPG-IPP Garching) and Fulvio Zonca (ENEA Frascati) for stimulating discussions. This manuscript is partly based upon work supported by the US Department of Energy, Office of Science, Office of Fusion Energy Sciences, and has been co-authored by Princeton University under Contract DE-AC02-09CH11466 with the US Department of Energy. The work was partly supported by the Early Career Research Program, project {\it Phase-Space Engineering of Supra-Thermal Particle Distribution for Optimizing Burning Plasma Scenarios}. The publisher, by accepting the article for publication, acknowledges that the United States Government retains a non-exclusive, paid-up, irrevocable, world-wide license to publish or reproduce the published form of this manuscript, or allow others to do so, for United States Government purposes.
\end{acknowledgments}

\appendix

%==============================================================================
\section{Forms of the resonance condition}
\label{apdx:res}

In an idealized tokamak that is perfectly symmetric in the toroidal angle $\zeta$, and whose magnetic field ${\bm B} = B\hat{\bm b}$ is independent of time, the guiding center (gc) motion of charged particles is confined to toroidal surfaces that are uniquely determined by a triplet of unperturbed constants of motion ${\bm C}$ \cite{HeidbrinkWhite20}. The guiding center position on such a drift orbit surface can be identified by poloidal ($\vartheta$) and toroidal ($\zeta$) angles, and the guiding center circulates the poloidal contour of its orbit surface with period
\begin{equation}
	\tau_{\rm pol}({\bm C}) \equiv \oint\frac{{\rm d}\vartheta}{\dot{\vartheta}_{\rm gc}} \quad \text{with} \quad \dot{\vartheta}_{\rm gc} \equiv \frac{{\rm d}\vartheta_{\rm gc}}{{\rm d}t}.
\end{equation}

\noindent Taking the poloidal transit time $\tau_{\rm pol}$ as a reference, one can define the mean angular frequencies
\begin{subequations}\vspace{-0.1cm}
	\begin{align}
		\omega_{\rm pol}({\bm C}) \equiv &\, \left<\smash{\dot{\vartheta}_{\rm gc}}\right>_{\rm pol} \equiv \frac{1}{\tau_{\rm pol}}\oint_{\tau_{\rm pol}}{\rm d}t\,\dot{\vartheta}_{\rm gc}(t) = \frac{2\pi}{\tau_{\rm pol}},
		\label{eq:wtrans_pol}
		\\
		\omega_{\rm tor}({\bm C}) \equiv &\, \left<\smash{\dot{\zeta}_{\rm gc}}\right>_{\rm pol} = \frac{\zeta_{\rm gc}(t+\tau_{\rm pol})- \zeta_{\rm gc}(t)}{\tau_{\rm pol}}.
		\label{eq:wtrans_tor}
	\end{align}
	\label{eq:wtrans}
\end{subequations}

\noindent Their ratio
\begin{equation}
	h({\bm C}) \equiv \frac{\omega_{\rm tor}}{\omega_{\rm pol}} \quad \stackrel{\rm drift}{\longleftarrow} \quad q(\Psi_{\rm P}) \equiv \frac{{\rm d}\Psi}{{\rm d}\Psi_{\rm P}},
	\label{eq:h}
\end{equation}

\noindent known as orbit helicity (or orbit pitch parameter) \cite{Shinohara18}, is the drift-kinetic generalization of the safety factor $q$ that measures magnetic field helicity, here written in terms of toroidal and poloidal magnetic fluxes, $\Psi$ and $\Psi_{\rm P}$.

When guiding centers on confined drift orbits (not intersecting the wall) interact with waves that
\begin{itemize}
	\item have low (sub-cyclotron) frequency $\omega \ll \omega_{\rm c}$,
	\item travel mainly along the magnetic field, like shear Alfv\'{e}n or acoustic waves (which satisfy the dispersion relation $\omega = k_\parallel v_{\rm A/S}$ with parallel wavenumber $k_\parallel = -i\hat{\bm b}\cdot\nablab$ and Alfv\'{e}n/sound speed $v_{\rm A/S}$),
	\item have low amplitudes, so that the unperturbed orbit surfaces ${\bm C}$ and transit frequencies (\ref{eq:wtrans}) are still good approximations,
\end{itemize}

\noindent the wave-particle resonance condition can be written as
\begin{equation}
	\omega = \underbrace{n\omega_{\rm tor}({\bm C}) - p\omega_{\rm pol}({\bm C})}\limits_{\equiv \Omega_{p,n}({\bm C})} \quad \text{with} \quad n,p \in \mathbb{Z}.
	\label{eq:res1}
\end{equation}\vspace{-0.25cm}

\noindent Eq.~(\ref{eq:res1}) is obtained by requiring that the phase velocity of a harmonic wave with toroidal and temporal dependence $\exp(in\zeta - i\omega t)$ matches the mean angular velocity of a charged particle's guiding center orbit. In the case of mirror-trapped particles, the integer $p$ is the bounce harmonic and only the toroidal wavelength enters Eq.~(\ref{eq:res1}) via the toroidal Fourier mode number $n$. In the case of passing particles, $p$ is the poloidicity of a resonance and incorporates primarily information about the mode structure in the poloidal angular direction $\vartheta$.

Note that $\omega_{\rm tor}$ contains contributions of rapid parallel streaming and slow magnetic drift. If one wishes to separate these two time scales, the resonance condition (\ref{eq:res1}) can be rewritten in the following form that appears, for instance, in Eq.~(27) of Ref.~\cite{Briguglio14} or Eq.~(4.171) of the review in Ref.~\cite{Chen16}:
\begin{equation}
	\omega = n\omega_{\rm D} + [(n\overline{q} - m)\overline{\sigma} + k]\omega_{\rm pol} \quad \text{with} \quad n,m,k \in \mathbb{Z}.
	\label{eq:res2}
\end{equation}

\noindent Here, $\overline{\sigma}$ identifies the orientation of the parallel guiding center velocity $u$ given by Eq.~(\ref{eq:u}), with $\overline{\sigma} = 0$ for mirror-trapped particles. $\overline{q}$ is some convenient measure for the mean field line helicity on the poloidal contour of the guiding center orbit.\footnote{In Ref.~\protect\cite{Briguglio14}, $\overline{q}$ is the poloidal-angle-averaged $q$. Alternatively, one may user the orbit-time integral $\left<q\right>_{\rm pol}$ as defined in Eq.~(\protect\ref{eq:wtrans_pol}).}
The first term in Eq.~(\ref{eq:res2}) approximately measures the slow component of $\omega_{\rm tor}$, which is known as the toroidal precession frequency,
\begin{equation}
	\omega_{\rm D} \equiv \omega_{\rm tor} - \overline{\sigma}\,\overline{q}\omega_{\rm pol} = \omega_{\rm tor} - \frac{\overline{q}}{h}\omega_{\rm tor} \approx \omega_{\rm tor} - \frac{\overline{u}_{\rm tor}}{R} \frac{\overline{q}}{h},
	\label{eq:wd}
\end{equation}

\noindent where we used $\overline{u}_{\rm tor} \equiv \overline{u}B_{\rm tor}/B \approx R\omega_{\rm tor}$ and $B \approx B_{\rm tor}$ to demonstrate how the subtracted term approximately measures rapid parallel streaming. Clearly, $\omega_{\rm D} \rightarrow 0$ vanishes in the absence of magnetic drift since $h \rightarrow q$ and $\overline{q}\rightarrow q$. The last term $k\omega_{\rm pol}$ in Eq.~(\ref{eq:res2}) is a redefined bounce harmonic with $k \equiv m\overline{\sigma} - p$, where $m$ is the wave field's main poloidal Fourier component with basis $e^{im\vartheta}$ in a certain metric of $\vartheta$. The rest of Eq.~(\ref{eq:res2}) absorbs the (canceling) counterpart $+n\overline{\sigma}\,\overline{q}\omega_{\rm pol}$ of the subtracted term in Eq.~(\ref{eq:wd}) and can then be recognized as an approximation of
\begin{align}
	-k_\parallel u =&\, \frac{n B^\zeta - m B^\vartheta}{B} u \approx \frac{nq - m}{q}\frac{u}{R}\\
 	\approx&\, \frac{1}{q}(nq-m)\omega_{\rm tor} = \sigma\frac{h}{q}(nq-m)\omega_{\rm pol},
\end{align}

\noindent where we used $q = B^\zeta/B^\vartheta = {\bm B}\cdot\nablab\zeta/{\bm B}\cdot\nablab\vartheta$ (in straight-field-line coordinates) and omitted the averages $\overline{(...)}$. One can now see that Eq.~(\ref{eq:res2}) approximately resembles
\begin{equation}
	0 \approx \omega + k_\parallel u - n\omega_{\rm D} - k \omega_{\rm pol}  \quad \text{with} \quad n,k \in \mathbb{Z},
	\label{eq:res0}
\end{equation}

\noindent that separates transit, precession and bounce resonances.
 
The arbitrariness of the integers $p$ in Eq.~(\ref{eq:res1}) and $k$ in Eq.~(\ref{eq:res2}) can be reduced by replacing the poloidal Fourier mode number $m$ in flux space by the dominant poloidal mode number $m_{\rm orb}^{(k)}$ of the mode projected onto the poloidal orbit contour \cite{Bierwage14}:
\begin{equation}
	p = m_{\rm orb}^{(k)}\sigma - k.
	\label{eq:res_meff}
\end{equation}

\noindent At least in the case of long-wavelength modes in the plasma core ($nq \sim \O(1)$), our experience has been that the dominant resonance for passing drift orbits often corresponds to $k=0$, or $k = \pm 1$ at the most. Larger values of $|k|$ tend to be inefficient (and, thus, irrelevant) higher-order resonances. This seems to be particularly true for long-wavelength Alfv\'{e}n modes interacting with near- or super-Alfv\'{e}nic fast ions, whose poloidal transit frequency $\omega_{\rm pol}$ is comparable to or even larger than the mode frequency $\omega$, so that neighboring resonances $k = \pm 1$ are far away, and may not even exist inside the plasma. This was also the case in the present study. In Fig.~\ref{fig:02_setup}, we had $m = 2$ but dominant $m_{\rm orb}^{(0)} = 1$. Indeed, the resonant vortices appearing in Figs.~\ref{fig:03_poink}--\ref{fig:06_scanf} all had poloidicity $p=1$.

In cases where amplitudes grow large and interaction times become short --- shorter than the period $\tau_{\rm pol}$ underlying the definitions in Eq.~(\ref{eq:wtrans}) --- one can generalize Eq.~(\ref{eq:res1}) by replacing the transit frequencies $\omega_{\rm tor}$ and $\omega_{\rm pol}$ with coefficients that measure toroidal and poloidal transit rates weighted by local fluctuation intensities:
\begin{equation}
	\omega = \left<\smash{n\dot{\zeta}_{\rm gc} - m_{\rm eff}\dot{\vartheta}_{\rm gc}}\right>_{\rm int} \quad \text{(local resonant kick)},
	\label{eq:res4}
\end{equation}

\noindent where $\left<...\right>_{\rm int}$ is a perturbation-amplitude-weighted average in the interaction region, and $m_{\rm eff}$ is an effective local poloidal mode number of the wave. Eq.~(\ref{eq:res4}) was previously used to explain enhanced prompt losses \cite{Zhang15,Heidbrink16}.

%\vspace{-0.7cm}
%==============================================================================
\setlength{\bibsep}{0.6pt}
\bibliographystyle{unsrt}
\bibliography{references}

\end{document}